\documentclass{aa}  

\usepackage{graphicx}
\usepackage{txfonts}
\usepackage{amsmath}    
\usepackage{xspace}
\usepackage{xcolor}
\usepackage[colorlinks=true,citecolor=teal]{hyperref}

\newcommand{\msun}{$M_\odot$\xspace}
\newcommand{\dd}{\text{d}}
\newcommand{\M}{\mathcal{M}}

\begin{document}

   \title{Likelihood of white dwarf binaries to dominate the astrophysical gravitational wave background in the mHz band}

\titlerunning{The WD Astrophysical GW Background}

   \author{Seppe Staelens
          \inst{1,2,3}\fnmsep\thanks{E-mail: ss3033@cam.ac.uk},
          \and
          Gijs Nelemans\inst{2, 1, 4}
          }

\authorrunning{S. Staelens \& G. Nelemans}

   \institute{Institute of Astronomy, KU Leuven, Celestijnenlaan 200D, 3001 Leuven, Belgium
         \and
             Department of Astrophysics/IMAPP, Radboud University, PO Box 9010,
6500 GL, The Netherlands
        \and
            DAMTP, Centre for Mathematical Sciences, University of Cambridge, Wilberforce Road, Cambridge CB3 0WA, United Kingdom
        \and
            SRON, Netherlands Institute for Space Research, Niels Bohrweg 4, 2333 CA Leiden, The Netherlands
             }

   \date{Received \today; accepted }

 
  \abstract
   {The astrophysical gravitational wave background (AGWB) is a collective signal of astrophysical gravitational wave sources dominated by compact binaries.  One key science goal of current and future gravitational wave detectors is to obtain its measurement. }
   {We aim to determine the population of compact binaries dominating the AGWB in the mHz band. We revisit and update an earlier work by \cite{2003MNRAS.346.1197F} to
model the astrophysical gravitational wave background sourced by extragalactic
white dwarf binaries in the mHz frequency band.}
   { We calculated the signal using a single-metallicity model for the white dwarf population in the Universe using the global star formation history.}
   {We estimate the white dwarf AGWB amplitude to be $\sim 60$\% higher than the earlier estimate. We also find that the overall shape of the white dwarf AGWB shows a good fit with a broken power law combined with an exponential cut-off.}
   {We compare our results to present-day best estimates for the background due to black hole and neutron star binaries, finding that the white dwarf component is likely to dominate in the mHz band. We provide an order-of-magnitude estimate that explains this hierarchy and we comment on the implications for future missions that aim to detect the AGWB. We also note that the black hole AGWB may only be detectable at high frequency. We outline several improvements that can be made to our estimate, however, these points are unlikely to change our main conclusion, which posits that the white dwarf AGWB dominates the mHz band. }


   \maketitle
%

\section{Introduction}

Apart from individual gravitational wave (GW) sources, such as the black hole (BH) and neutron star (NS) mergers detected by the LIGO/Virgo/Kagra (LVK) collaboration \citep{2021arXiv211103606T} in the relatively nearby Universe, there is an isotropic stochastic signal coming from all GW-emitting sources in the Universe \citep[e.g.][]{2010CQGra..27s4007S,2019RPPh...82a6903C}. In particular, all compact binaries that are evolving due to the emission of GW form a signal that displays a characteristic spectral shape \citep{2001astro.ph..8028P} and this is generally referred to as the astrophysical GW background \citep[denoted as AGWB to distinguish it from potential stochastic GW signals originating in the Early Universe, e.g.][]{2001MNRAS.324..797S,2003MNRAS.346.1197F,2015A&A...574A..58K, 2023LRR....26....2A}. It is a broadband signal that extends from the high-frequency band covered by the LVK detectors to the mHz band, which will be covered by future space GW detectors such as LISA \citep{2017arXiv170200786A}, Taiji \citep{2021PTEP.2021eA108L}, and TianQin \citep{2016CQGra..33c5010L}. 

Although the AGWB is a collective signal, it provides information on the global properties of the population of compact binaries that are not individually detectable because they are too far away for their intrinsic luminosity. Following the discovery that the binary BH (BBH) population is both greater and contains more massive BH than previously believed \citep[see][]{2016ApJ...818L..22A}, several investigations have calculated the expected AGWB caused by BBHs \citep[and binary neutrons stars, BNSs, e.g.][]{2016MNRAS.461.3877D,2022A&A...660A..26B,2023JCAP...08..034B,2023MNRAS.tmp.2801L}. These studies concluded that it would be detectable by the LISA mission and future ground based detectors such as Einstein Telescope \citep{2010CQGra..27s4002P} and Cosmic Explorer \citep{2019BAAS...51g..35R}. 

A more detailed understanding of the AGWB is also necessary to fully characterise and remove it from the data to uncover underlying signals, such as GWs from the Early Universe, which have become an increasingly popular research topic \citep[see e.g.][]{2009JCAP...12..024C,2012JCAP...06..027B,2023LRR....26....5A}, or GWs from faint individual sources.

In the mHz band, apart from BH and NS binaries, there is also a contribution from white dwarf (WD) and other binaries. \citet{2003MNRAS.346.1197F} performed a detailed calculation of the AGWB excluding BH and NS systems but did not discuss in detail the notion of whether it would be detectable by the LISA mission. Meanwhile, \citet{2001MNRAS.324..797S} concluded that the WD AGWB would dominate. We note that due to the much larger size of WD and other stars, they do not contribute to the AGWB above frequencies of about 100 mHz \citep{2003MNRAS.346.1197F}. We also stress that this WD AGWB should not be confused with the stochastic signal of unresolved WD binaries within our Milky Way, referred to as the GW 'foreground'; although the latter is stochastic, it is far from isotropic \citep[e.g.][]{2005PhRvD..71l2003E}. Given the conclusion that the BH AGWB can (easily) be detected by the LISA mission \citep[e.g.][]{2016JCAP...04..001C,2023JCAP...08..034B}, we find that it is time to reinvestigate the relative importance of the WD AGWB.

In this paper, we compare the BH AGWB as derived from the LVK measurements with a new estimate of the WD AGWB. In Section \ref{methods}, we describe our methods. In Section \ref{results}, we present our results and in Section \ref{discussion}, we discuss these results and their limitations, followed by our conclusions. We use a cosmology based on the parameters from \citet{planck_collaboration_planck_2020}.

\section{Methods}\label{methods}
The AGWB is described in terms of the dimensionless energy density spectrum:
\begin{equation}\label{eq: def Omega 1}
    \Omega_{\text{GW}}(f_r) = \frac{1}{\rho_c c^2}\frac{\dd \mathcal{E}_{\text{GW}}}{\dd \ln f_r}\,,
\end{equation}
where $\rho_c = \frac{3H_0^2}{8\pi G}$ is the critical density of the Universe, $\mathcal{E}_{\text{GW}}$ is the present-day energy density in GWs and $f_r$ is the received GW frequency. 
In an isotropic and homogeneous Universe, the present-day GW energy density due to a population of compact binaries in the inspiral phase is given by \citep{phinney_practical_2001}
\begin{align}
    \mathcal{E}_{\text{GW}}
    & =  \int_0^\infty \frac{\dd f_r}{f_r} f_r \int_0^\infty \dd z \,N(z)\frac{\dd E_{\text{GW}}}{\dd f_e} \label{eq: rho gw}\,.
\end{align}
In this expression, $N(z)$ is the number density of sources with energy spectrum $\frac{\dd E_{\text{GW}}}{\dd f_e}$. The latter can be approximated by the leading quadrupole order of the emitted GW radiation \citep{hawking1987three}
\begin{equation}
    \frac{\dd E_{\text{GW}}}{\dd f_e} = \frac{\pi^{2/3}}{3}G^{2/3}\mathcal{M}^{5/3}f_e^{-1/3}\qquad \text{ for } f_{\text{min}} < f_e < f_{\text{max}}\,.
\end{equation}
The frequency of the GW is related to the orbital frequency $\nu$ as $f_e = 2\nu$. We note that $f_e$ is the GW frequency in the rest frame of the emitting source, which is redshifted as $f_e = f_r(1+z)$ by the time the signal is detected. 
The limiting frequencies depend on the source properties: the lower limit, $f_{\text{min}}$, is set by the separation of the system after circularisation, and is such that the system changes frequency significantly within a Hubble time. The frequency at which the source emits increases over time as \citep{2003MNRAS.346.1197F}:
\begin{equation}\label{eq: nu dot}
    \dot{\nu} \equiv \frac{\dd \nu}{\dd t} = K \nu^{11/3}\,,
\end{equation}
until the binary components come into Roche lobe contact or the evolution becomes dominated by tidal forces. This sets a minimal orbital separation, and therefore a maximal frequency $f_{\text{max}}$, for the binary. Equation (\ref{eq: nu dot}) can be integrated as follows:
\begin{equation}\label{eq: nu t}
    \nu(t)^{-8/3} - \nu_0^{-8/3} = \frac{8K(t_0-t)}{3}\,,
\end{equation}
where $t_0$ and $\nu_0$ are the time and frequency of the binary at birth, respectively. The constant in Eq. (\ref{eq: nu dot}) depends on the chirp mass as
follows:\ \begin{equation}\label{eq: K}
    K = \frac{96}{5}\left(2\pi\right)^{8/3}\left(\frac{G \M}{c^3}\right)^{5/3} \approx 3.7 \cdot 10^{-6} \left(\frac{\M}{M_\odot}\right)^{5/3} \text{ s}^{5/3}\,.
\end{equation}
If we allow the population to have varying chirp mass, the background is given by \citep{2003MNRAS.346.1197F, renzini_stochastic_2022}:
\begin{equation}\label{eq: f23}
    \Omega_{\text{GW}}^{\text{CB}}(f_r) = \frac{(\pi G f_r)^{2/3}}{3\rho_c c^2}\int_0^\infty \dd z \frac{1}{(1+z)^{1/3}} \int_{\M_{\text{min}}}^{\M_{\text{max}}} \dd\M\frac{\dd N}{\dd \M} \M^{5/3} \,.
\end{equation}
Therefore, if the integrals do not have a frequency dependence, the expected spectral dependence of the AGWB sourced by compact binaries is:
\begin{equation}\label{eq: f23 ref}
    \Omega \left(f\right) = \Omega\left(f_{\text{ref}}\right) \left(\frac{f}{f_{\text{ref}}}\right)^{2/3}\,,
\end{equation}
where $f_{\text{ref}}$ is a reference frequency to which the background is normalised.

\subsection{Estimating the AGWB}\label{estimate} 

This section aims to provide an estimate of the different contributions to the compact binary AGWB in the LISA frequency band. We first focus on the contributions of the BBH and BNS signals, leaving out the NS+BH population for simplicity (and it is subdominant). These contributions are obtained by extrapolating the current best estimates in the LVK band using Eq. (\ref{eq: f23 ref}). In reality, the signal may drop off somewhat more steeply, especially at low frequencies \citep[e.g.][]{2001MNRAS.324..797S}, so this extrapolation is an upper bound. The current upper limit on the AGWB in the LVK band is \citep{PhysRevD.104.022004}:
\begin{equation}
\Omega_{\text{GW}}(25 \text{ Hz}) \leq 3.4 \cdot 10^{-9}\,
,\end{equation}
at the $95\%$ confidence interval. The best estimates, based on current knowledge of the compact binary population are \citep{2023PhRvX..13a1048A}: 
\begin{align}
    \Omega_{\text{BBH}}(25 \text{ Hz}) & = 5.0^{+1.4}_{-1.8}\cdot 10^{-10}\,, \\
    \Omega_{\text{BNS}}(25 \text{ Hz}) & = 0.6^{+1.7}_{-0.5}\cdot 10^{-10}\,.
\end{align}
The WD contribution is obtained by updating the work of \citet{2003MNRAS.346.1197F}. The method we followed is based on their Sec. 6. It starts from an alternative formulation of Eq. (\ref{eq: def Omega 1}):
\begin{equation}\label{eq:Omega_indiv}
    \Omega(f_r) = \frac{1}{\rho_c c^3} f_r F_{f_r}\,,
\end{equation}
where $F_{f_r} = \frac{\dd F(f_r)}{\dd f_r}$ is the specific flux received from an object with specific luminosity, $L_{f_e}$, at redshift, $z$. 
It is given by:
\begin{equation}\label{eq: spec flux}
    F_{f_r} = \frac{L_{f_e}}{4\pi d_L(z)^2}\left(\frac{\dd f_e}{\dd f_r}\right)\,,
\end{equation}
where $d_L = (1+z)d_M$ is the luminosity distance to redshift, $z$, and $d_M$ is the proper motion distance. 
In the case of a large number of systems, 
all these contributions can be added together by introducing the specific luminosity density, $l_{f_e}$, defined as $\dd L_{f_e} (z) = l_{f_e}(z) \dd V(z)$. Then,
$\dd V(z)$ is the comoving volume element at redshift $z$, namely, $\dd V(z) = 4\pi d_M(z)^2 \dd \chi$, with $\chi(z)$ the proper motion distance.
The specific flux given in Eq. (\ref{eq: spec flux}) can then be rewritten as:
\begin{equation}\label{eq: spec flux int z}
    F_{f_r} = \int_{z=0}^\infty \frac{l_{f_e}(z)}{(1+z)^2 }\left(\frac{\dd f_e}{\dd f_r}\right) \dd \chi(z)\,.
\end{equation}
In order to evaluate this numerically, the
integral over the redshift is discretized\footnote{Note: \cite{2003MNRAS.346.1197F} discretized the integral over cosmic time: $\dd \chi = - (1+z)\,c\,\dd T$. We have checked that our method is in agreement with the latter by calculating the AGWB using both discretizations. Furthermore, we have checked that increasing the number of redshift bins does not significantly alter the result.} in 20 bins over the interval of $z \in [0,8]$. Furthermore, the frequency range $[10^{-5}, 1]$ Hz is divided into 50 bins, and the received flux per bin is calculated as:
\begin{align}
    F_{f_{r_1} \to f_{r_2}} & = \sum_i  \int_{f_{r_1}(1+z_i)}^{f_{r_2}(1+z_i)} \frac{l_{f_e}}{(1+z_i)^2 }\dd f_e \Delta \chi(z_i)\,, \label{eq: flux bin}
\end{align}
where we integrated over $f_r$ and performed a change of variables as $f_r \to f_e$. This expression essentially shows that we integrate over all the source frequencies that get redshifted to the correct observed frequency bin. 
The specific luminosity density is determined as the sum of the contributions of different systems, labeled as $k$, and equal to
\begin{equation}\label{eq: spec lum density}
    l_{f_e} = \sum_k L_k(f_e) n_k(f_e, z)\,.
\end{equation}
The luminosity is given by \citep{peters_gravitational_1963}
\begin{equation}\label{eq: spec lum}
    L_k(f_e) = \frac{32\pi^{10/3}}{5}\frac{G^{7/3}}{c^5}\M_k^{10/3}f_e^{10/3}\,,
\end{equation}
and $n_k(f_e, z)$ is the specific number density of systems of type $k$ at redshift $z$, emitting GWs at a frequency $f_e$. The specific number density scales as $n_k(f) = A_k f^{-11/3}$ (see Eq. \ref{eq: number density}), where the proportionality constant $A_k$ is such that: 
\begin{equation}
    \int_{\text{bin}} A_k f ^{-11/3} \dd f = n_{\text{bin}}(k;z)\,,
\end{equation}
with $n_{\text{bin}}(k;z)$ the number density of systems $k$ at redshift $z$ emitting GWs with frequencies in the correct bin. The latter can be estimated by multiplying the star formation rate per unit of volume, $\psi,$ with the time it takes a binary to traverse the frequency bin:
\begin{equation}\label{eq: n bin}
    n_{\text{bin}}(k;z) \approx \mathcal{P}_k\cdot \psi(z+\Delta z)\cdot \Delta t(k;\text{bin})\,.
\end{equation}
The factor $\mathcal{P}_k$ reflects the proportion of the formed stars that end up in a binary with properties corresponding to label $k$. The value of $\Delta z$ reflects an additional redshift to reflect the conditions at formation of the binary, before evolving to the frequency at the lower end of the bin, which happens at redshift $z$. This time delay, as well as $\Delta t$ can be calculated using Eq. (\ref{eq: nu t}).\\
The population model and star formation rate history that we used are explained in the next section. Finally, the energy density in the frequency bin is calculated as
\begin{equation}
    \Omega(f_r) \approx \frac{1}{\rho_c c^3} \frac{f_r F_{f_{r_1}\to f_{r_2}}}{f_{r_2} - f_{r_1}}\,,
\end{equation}
where $f_r$ is now an appropriately chosen frequency that represents the bin.

\subsection{The WD population model and star formation rate history}\label{wd_model}

\begin{figure}
    \includegraphics[width=\columnwidth]{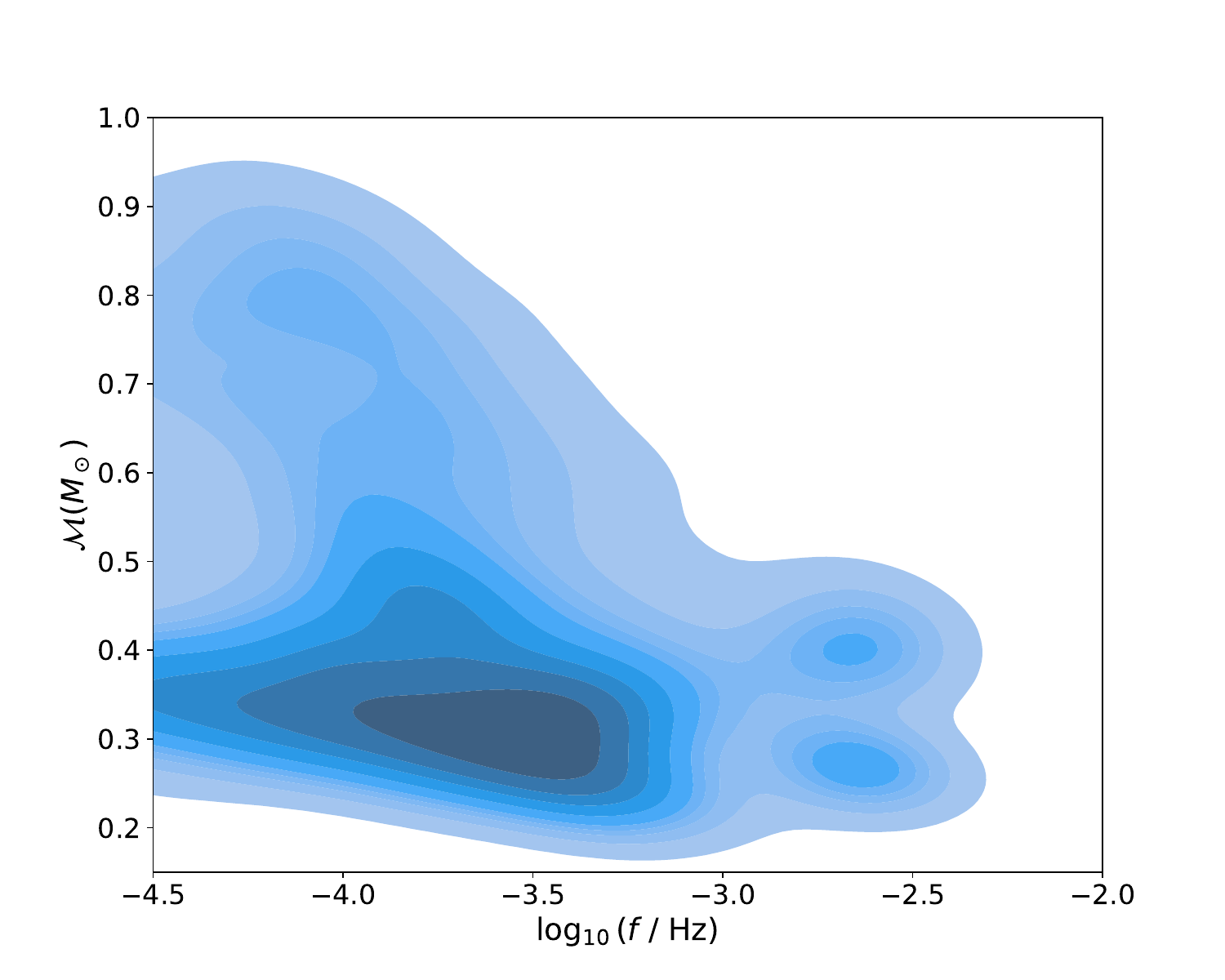}
\caption{Density plot (linear scale) of the initial properties of the WD population: chirp mass, $\mathcal{M,}$ versus GW frequency at the time of formation.}
\label{fig:wd_pop}
\end{figure}

In order to sample the population of WD in the different redshift shells, we calculated a model for the population of double WD based on the SeBa population synthesis code \citep{1996A&A...309..179P,2001A&A...365..491N,2012A&A...546A..70T}. This code also has also been used to make synthetic models for the population of double WD detectable by LISA in the Galaxy and nearby galaxies \citep{2001A&A...375..890N,2020A&A...638A.153K,2023MNRAS.521.1088K} and used in the LISA Data Challenges \citep{2022arXiv220412142B}. 
For this first estimate of the WD AGWB, we used a single (solar) metallicity and simulate 250,000 binaries with initial masses between 0.9 and 11 \msun, distributed according to the initial mass function (IMF) of \citet{2001MNRAS.322..231K}, with a flat mass ratio distribution and initial separations flat in $\log a$. For each of the systems, the evolution was followed and we selected 14,418 systems that form a close ($a < 500 R_\odot)$ double WD. 
The properties with which the WD binaries are formed (chirp mass and GW frequency) are shown in Figure~\ref{fig:wd_pop}, with the majority of sources having low chirp mass and frequencies at formation below 1 mHz, with an additional tail of systems forming with frequencies above 1 mHz.

The population we obtained corresponds to $4\cdot 10^6 M_\odot$ of total star formation, taking the complete IMF and the mass dependent binary fraction \citep{2017ApJS..230...15M} into account. We then take this population to be representative for every redshift shell in Eq. (\ref{eq: flux bin}). This means that we sum over all the binaries of our population in Eq. (\ref{eq: spec lum density}), and the factor $\mathcal{P}_k$ in Eq. (\ref{eq: n bin}) can be replaced as follows:\ 
\begin{equation}
    n_{\text{bin}}(k;z) \approx \frac{\psi(z+\Delta z)}{4\cdot 10^6 M_\odot} \cdot \Delta t(k;\text{bin})\,.
\end{equation}
The star formation history is taken to be the one in \citet{madau_cosmic_2014}:
\begin{equation}\label{eq: SFR1}
    \psi(z) = 0.015\frac{(1+z)^{2.7}}{1+[(1+z)/2.9]^{5.6}} \quad M_\odot \text{ yr}^{-1} \text{ Mpc}^{-3}\,.
\end{equation}
This exhibits a peak approximately 3.5 Gyr after the Big Bang, at $z \approx 1.9$.

\section{Results}\label{results}

\subsection{Order of magnitude}

Before we share the results of the estimate of the AGWB, we can make an order-of-magnitude estimate of the relative importance of WD, NS, and BH to the AGWB, by comparing the number density of sources as function of frequency of each of the classes and their GW luminosity. The number density of sources can easily be estimated from Eq. (\ref{eq: nu dot}) via
\begin{equation}\label{eq: number density}
    \frac{dN}{df} = \frac{dN}{dt}\frac{dt}{df} = \mathcal{R}/\dot{f} = \mathcal{R} \mathcal{M}^{-5/3} f^{-11/3}
,\end{equation}
with $\mathcal{R}$ as the rate at which systems arrive at a certain frequency. For sources that are inspiralling, $\mathcal{R}$ can be approximated by the merger rate of the sources. The contribution to the AGWB of two sets of sources with such density is simply this number density times the signal per source (see Eq. \ref{eq:Omega_indiv}):
\begin{equation}
    \Omega_{\rm GW,i} \propto   \frac{dN}{df} f F_{\rm GW} = \mathcal{R} \mathcal{M}^{5/3} f^{2/3}
,\end{equation}
where we use Eqs. (\ref{eq: spec flux}) and (\ref{eq: spec lum}) to estimate the flux (assuming they have the same distance distribution).
\begin{table}
        \centering
        \caption{Order-of-magnitude estimates of the AGWB using the merger rate per Milky Way equivalent galaxy (MWEG) and a typical chirp mass of WD, BH, and NS binaries.}
        \label{tab:orderofmag}
        \begin{tabular}{lccr} 
                \hline
                source & $\mathcal{R}$ & $\mathcal{M}$  & $\mathcal{R} \mathcal{M}^{5/3}$ \\
                   &  (MWEG yr)${}^{-1}$ & (\msun) & relative \\
                \hline
                WD & $1 \times 10^{-2}$ & $0.3$ &  $1$ \\
                BH & $3 \times 10^{-6}$ & $20$ &   $0.4$\\
                NS & $1 \times 10^{-5}$ & $1.2$ & $0.01$\\
                \hline
        \end{tabular}
\end{table}

We assumed a typical chirp mass for BBH of 20 \msun, for BNS  of 1.2 \msun and for WD binaries of 0.3 \msun (components of 0.3 and 0.4 \msun). For BBH and BNS, we estimated the merger rates in the Milky Way by scaling from the volume rates of the LVK collaboration \citep{2023PhRvX..13a1048A}, as in \citet{2010CQGra..27q3001A}. 
We used the same method as in \citet{2004MNRAS.349..181N} to obtain the merger rate of the WD population described above in the Milky Way. This results in the values given in Table~\ref{tab:orderofmag}, demonstrating that the WD background is likely to dominate, with the BH background reaching the same order of magnitude. The NS background is likely to be substantially less important. 

\subsection{The estimated AGWB}
\begin{figure}
        \includegraphics[width=\columnwidth]{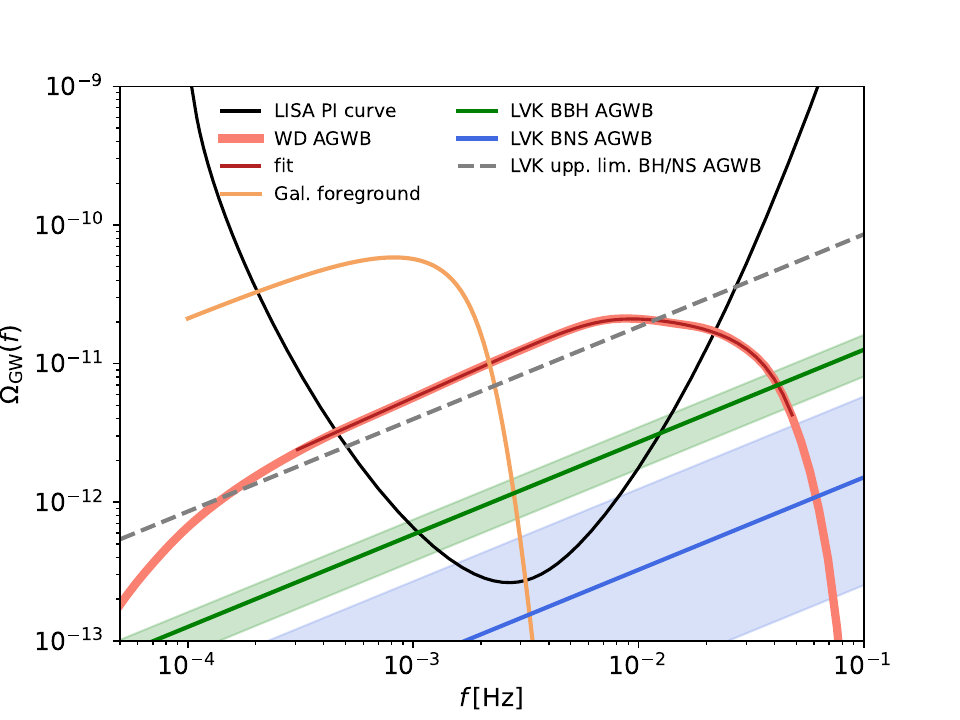}
    \caption{Resulting WD component (thick salmon) of the AGWB compared to the LVK results (upper limit to BH/NS AGWB (dashed grey) and estimates for the BBH and BNS components in green and blue), the LISA Powerlaw Integrated sensitivity \citep[black][]{2013PhRvD..88l4032T,2020PhRvD.101l4048A}, and an estimate of the Galactic foreground (brown) based on \citet{2021PhRvD.104d3019K}. A fit to the WD component is shown in dark red.}
   \label{fig:AGWB}
\end{figure}

The resulting components of the AGWB in the mHz frequency band calculated according to the methods discussed above are shown in Figure~\ref{fig:AGWB}, with the LVK extrapolations in green (BBH) and blue (BNS) and the calculated WD signals in orange. It is seen that the calculated WD component is significantly larger than the BH (and NS) component - about an order of magnitude. Furthermore, it is also seen to be larger than the extrapolated upper limit in the LVK frequency band, and well above the LISA sensitivity curve in the range $10^{-3} - 10^{-2}$ Hz.

The WD component displays a turnover around 10 mHz and the BH component becomes dominant around 40 mHz. This can be explained due to the fact that most WD binaries have $f_{\text{max}} \lesssim 100 $ mHz, causing a steady decline in number of sources that can contribute to the total background signal. Below $\sim$0.5 mHz the WD signal also drops off, due to the fact that most of the systems form in that region and also because the merger time for the systems becomes larger than the Hubble time \citep[see also][]{2003MNRAS.346.1197F}.

\begin{figure}
        \includegraphics[width=\columnwidth]{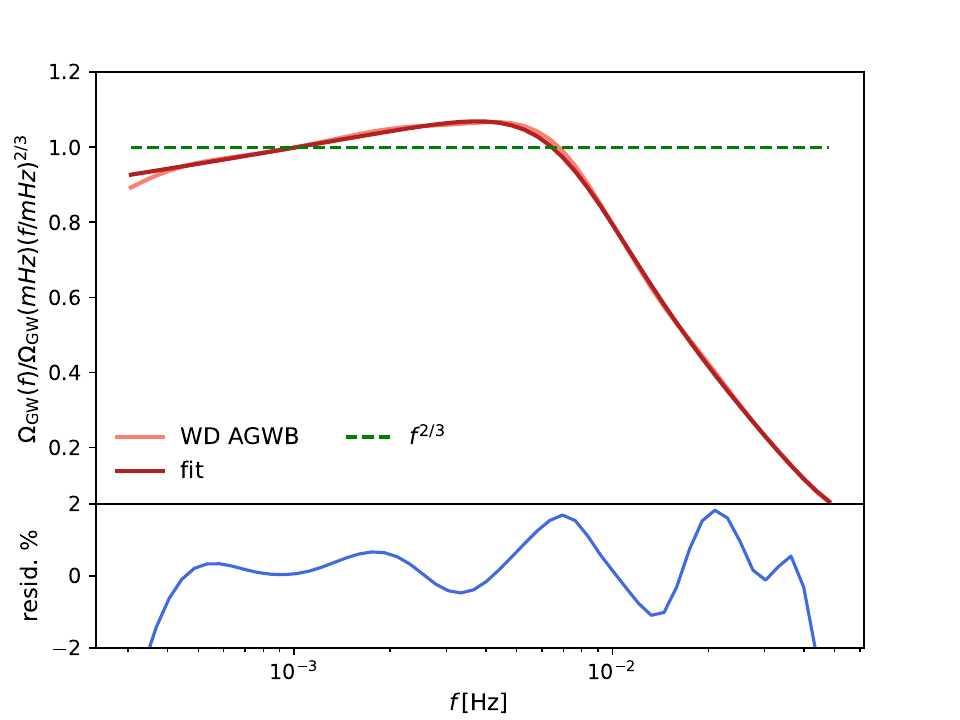}
    \caption{Comparison of the WD AGWB (salmon) and the fit (\ref{eq: fit}) (dark red) to a purely $f^{2/3}$ signal (green dashed) by dividing the curves by $f^{2/3}$. The WD AGWB deviates from the expected  $2/3$  slope, but the residuals (bottom) between the WD AGWB and the fit are below 2\%.}
   \label{fig:fit}
\end{figure}

The WD signal seems to follow a straight power law. In order to investigate this further, we plot in Figure~\ref{fig:fit} the signal in the region above 0.3 mHz divided by a pure $f^{2/3}$ power law normalised at 1 mHz. It is clear that there is a small deviation from a straight power law and also that the slope is slightly steeper than the expected $2/3$ (Eq. \ref{eq: f23 ref}). 
The signal is fitted well by a broken power law with an exponential cutoff:
\begin{align}
    \Omega_{\rm GW}(f) = & A\, \left(\frac{f}{\hat{f}}\right)^{0.73}  \left[1 + \left(\frac{f}{\hat{f}}\right)^{4.1}\right]^{-0.23} \cdot \exp\left(-B f^3\right) \label{eq: fit}
\end{align}
where $ A= 2.4\times 10^{-11}, B =  1.2\times 10^4$ and $\hat{f} = 7.2$ mHz.
The bottom panel of Figure~\ref{fig:fit} shows the residuals of the signal with respect to the fit, which for this simple fit remain below 2 percent. 
So, this fit is a good enough approximation for most studies.

\begin{figure}
        \includegraphics[width=\columnwidth]{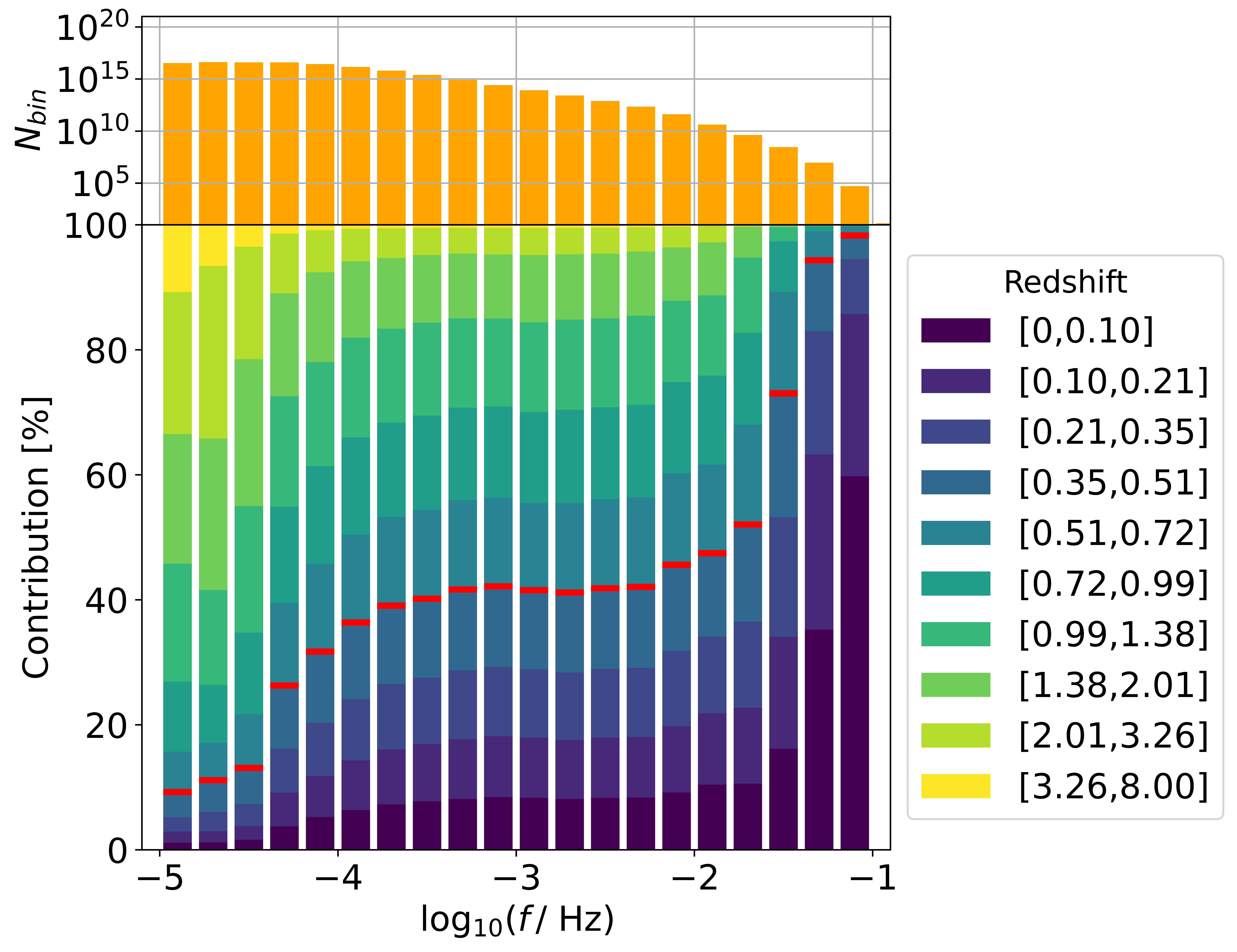}
    \caption{Contribution of different redshift shells in Eq. (\ref{eq: flux bin}) to the WD component of the AGWB for 20 frequency bins in the range $[10^{-5}, 0.1]$ Hz (bottom). 
    The red lines indicate the total contribution of the $z \le 0.5$ Universe.
    The high redshift contributions are small for the simple model we use and overall the signal is dominated by the $z \le 2$ Universe. The number of double WD, summed over redshift, that source the AGWB for the different frequency bins (top).}
   \label{fig:redshift_contributions}
\end{figure}

Finally, we note that the WD AGWB is dominated by the $z \leq 2$ Universe, as one could naively expect since the signals of distant WD are relatively weak. This is shown in Figure~\ref{fig:redshift_contributions} and justifies the fact that we only used redshift bins in the range $z \in [0,8]$.
This figure also shows that the signal at the highest frequencies is fully determined by the $z \lesssim 0.5$ Universe: this is due to the fact that very few systems merge in this frequency bin and any redshift pushes the merger to lower frequency bins. The pushing of the frequency to lower bins by the redshift explains why the relative contribution of the nearby Universe drops as the frequency decreases. 
Figure \ref{fig:redshift_contributions} also shows the number of double WDs that contribute at different frequencies: results are shown for the case where the frequency range $[10^{-5}, 0.1]$ Hz is divided into 20 bins, for the sake of clarity.
Adding everything up, we find a total of $\sim 1.5 \cdot 10^{17}$ binary systems that make up the background as calculated here.

\section{Discussion and conclusions}\label{discussion}

Given the amplitude of the WD component in Figure \ref{fig:AGWB}, it is expected that it can be very well measured by LISA. Furthermore, the relative amplitudes show that, if LISA detects an AGWB signal in the mHz regime, it is likely dominated by the WDs. This means that it is likely hard to make statements about the BH (and NS) population based on a measurement of the AGWB unless there is a way to disentangle the two, or to detect the high-frequency component of the AGWB above 40 mHz. 

In comparison to \citet{2003MNRAS.346.1197F}, the computation presented in this work results in a higher signal. They found $\Omega_{\text{WD}} (1\text{ mHz}) = 3.57 \cdot 10^{-12}$, whereas we find $\Omega_{\text{WD}} (1\text{ mHz}) = 5.7 \cdot 10^{-12}$.
There are several factors that contribute to this difference. First, they use a star formation history that is about a factor of 2 lower than what we use, at least at the low redshifts that dominate the background. Secondly, their population synthesis code is different, which will lead to different results. As an indication, \citet{2003MNRAS.346.1197F} find a local WD space density of $9 \cdot 10^{-3} \text{pc}^{-3}$, whereas \citet{2001A&A...365..491N} - also using the SeBa code -- find $19 \cdot 10^{-3} \text{pc}^{-3}$. 
However, since 2001 the SeBa code has been significantly updated, so this comparison should not be taken as a specific value, but as an indication that a difference in normalisation of a factor $\sim 2$ between codes is not surprising.  \citet{2003MNRAS.346.1197F} find a range of reference values between $1\cdot 10^{-12} \lesssim \Omega_{\text{WD}} (1\text{ mHz}) \lesssim 6\cdot 10^{-12} $ in their models. Therefore, the value found in this work lies at the upper end of this range. We defer an in depth investigation of the variation of the shape and normalisation of the WD AGWB to future work. 

Our estimate of the WD AGWB can be improved and extended in several ways, although this will not change our main conclusion that the WD AGWB likely dominates over the BH AGWB. First, we can do a study that takes the metallicity of the star formation \citep[e.g.][]{2019MNRAS.488.5300C} into account as done with SeBa before \citep[e.g.][]{2014A&A...569A..42V,2020A&A...638A.153K}, since metallicity can change the binary evolution. Additionally, there is also more and more evidence that metallicity also changes the initial binary parameters 
\citep{2018ApJ...854..147B,2019ApJ...875...61M}. In addition, there are many uncertainties in binary evolution so a study varying these will give an indication of the astrophysical uncertainties in the model. 

A more detailed model could also look into deviations from isotropy, since a significant fraction of the signal is found to originate in the nearby Universe, where the assumption of isotropy is not completely correct. This significantly complicates the modelling, but several studies showed that anisotropies could potentially be detected and contain information on the source population \citep[e.g.][]{2020MNRAS.493L...1C}. More detailed modelling will also offer more insight into the expected deviations of the signal from simple power laws and the shape of the turn over. This then provides input to detailed studies on the detectability of these features, for instance, with LISA. Finally, we note here that we neglected several types of binaries that will also contribute to the AGWB, albeit at a lower level than in the case of the double WD, such as interacting double WD binaries (AM CVn stars) and binaries containing helium stars or low-mass main sequence stars. 

In conclusion, we simulated the WD AGWB using relatively little, but likely sufficient, detail to show that it will dominate over the AGWB from BH and NS binaries in the mHz regime. This offers an opportunity to study the WD binary population to much larger distances, while hampering the detection of the BH AGWB with missions such as LISA. The WD signal reaches a peak around 10 mHz and at higher frequencies the BH AGWB will become the dominant signal. The detectability of this transition by LISA and other mHz missions ought to be studied in detail.

\begin{acknowledgements}
We thank the referee for valuable suggestions and comments.
Furthermore, we want to thank Sophie Hofman for pointing out errors in the initial code we used.
G.N. is supported by the Dutch science foundation NWO. S.S. is supported by the Centre for Doctoral Training (CDT) at the University of Cambridge funded through STFC.
\end{acknowledgements}

\bibliographystyle{aa} 
\bibliography{WD_AGWB} 

\begin{thebibliography}{50}
\expandafter\ifx\csname natexlab\endcsname\relax\def\natexlab#1{#1}\fi

\bibitem[{{Abadie} {et~al.}(2010){Abadie}, {Abbott}, {Abbott}, {Abernathy},
  {Accadia}, {Acernese}, {Adams}, {Adhikari}, {Ajith}, {Allen}, {Allen},
  {Amador Ceron}, {Amin}, {Anderson}, {Anderson}, {Antonucci}, {Aoudia},
  {Arain}, {Araya}, {Aronsson}, {Arun}, {Aso}, {Aston}, {Astone}, {Atkinson},
  {Aufmuth}, {Aulbert}, {Babak}, {Baker}, {Ballardin}, {Ballmer}, {Barker},
  {Barnum}, {Barone}, {Barr}, {Barriga}, {Barsotti}, {Barsuglia}, {Barton},
  {Bartos}, {Bassiri}, {Bastarrika}, {Bauchrowitz}, {Bauer}, {Behnke}, {Beker},
  {Belczynski}, {Benacquista}, {Bertolini}, {Betzwieser}, {Beveridge},
  {Beyersdorf}, {Bigotta}, {Bilenko}, {Billingsley}, {Birch}, {Birindelli},
  {Biswas}, {Bitossi}, {Bizouard}, {Black}, {Blackburn}, {Blackburn}, {Blair},
  {Bland}, {Blom}, {Blomberg}, {Boccara}, {Bock}, {Bodiya}, {Bondarescu},
  {Bondu}, {Bonelli}, {Bork}, {Born}, {Bose}, {Bosi}, {Boyle}, {Braccini},
  {Bradaschia}, {Brady}, {Braginsky}, {Brau}, {Breyer}, {Bridges}, {Brillet},
  {Brinkmann}, {Brisson}, {Britzger}, {Brooks}, {Brown}, {Budzy{\'n}ski},
  {Bulik}, {Bulten}, {Buonanno}, {Burguet-Castell}, {Burmeister}, {Buskulic},
  {Byer}, {Cadonati}, {Cagnoli}, {Calloni}, {Camp}, {Campagna}, {Campsie},
  {Cannizzo}, {Cannon}, {Canuel}, {Cao}, {Capano}, {Carbognani}, {Caride},
  {Caudill}, {Cavagli{\`a}}, {Cavalier}, {Cavalieri}, {Cella}, {Cepeda},
  {Cesarini}, {Chalermsongsak}, {Chalkley}, {Charlton}, {Chassande Mottin},
  {Chelkowski}, {Chen}, {Chincarini}, {Christensen}, {Chua}, {Chung}, {Clark},
  {Clark}, {Clayton}, {Cleva}, {Coccia}, {Colacino}, {Colas}, {Colla},
  {Colombini}, {Conte}, {Cook}, {Corbitt}, {Corda}, {Cornish}, {Corsi},
  {Costa}, {Coulon}, {Coward}, {Coyne}, {Creighton}, {Creighton}, {Cruise},
  {Culter}, {Cumming}, {Cunningham}, {Cuoco}, {Dahl}, {Danilishin},
  {Dannenberg}, {D'Antonio}, {Danzmann}, {Dari}, {Das}, {Dattilo}, {Daudert},
  {Davier}, {Davies}, {Davis}, {Daw}, {Day}, {Dayanga}, {De Rosa}, {DeBra},
  {Degallaix}, {del Prete}, {Dergachev}, {DeRosa}, {DeSalvo}, {Devanka},
  {Dhurandhar}, {Di Fiore}, {Di Lieto}, {Di Palma}, {Emilio}, {Di Virgilio},
  {D{\'\i}az}, {Dietz}, {Donovan}, {Dooley}, {Doomes}, {Dorsher}, {Douglas},
  {Drago}, {Drever}, {Driggers}, {Dueck}, {Dumas}, {Eberle}, {Edgar},
  {Edwards}, {Effler}, {Ehrens}, {Engel}, {Etzel}, {Evans}, {Evans}, {Fafone},
  {Fairhurst}, {Fan}, {Farr}, {Fazi}, {Fehrmann}, {Feldbaum}, {Ferrante},
  {Fidecaro}, {Finn}, {Fiori}, {Flaminio}, {Flanigan}, {Flasch}, {Foley},
  {Forrest}, {Forsi}, {Fotopoulos}, {Fournier}, {Franc}, {Frasca}, {Frasconi},
  {Frede}, {Frei}, {Frei}, {Freise}, {Frey}, {Fricke}, {Friedrich},
  {Fritschel}, {Frolov}, {Fulda}, {Fyffe}, {Gammaitoni}, {Garofoli}, {Garufi},
  {Gemme}, {Genin}, {Gennai}, {Gholami}, {Ghosh}, {Giaime}, {Giampanis},
  {Giardina}, {Giazotto}, {Gill}, {Goetz}, {Goggin}, {Gonz{\'a}lez},
  {Gorodetsky}, {Go{\ss}ler}, {Gouaty}, {Graef}, {Granata}, {Grant}, {Gras},
  {Gray}, {Greenhalgh}, {Gretarsson}, {Greverie}, {Grosso}, {Grote},
  {Grunewald}, {Guidi}, {Gustafson}, {Gustafson}, {Hage}, {Hall}, {Hallam},
  {Hammer}, {Hammond}, {Hanks}, {Hanna}, {Hanson}, {Harms}, {Harry}, {Harry},
  {Harstad}, {Haughian}, {Hayama}, {Heefner}, {Heitmann}, {Hello}, {Heng},
  {Heptonstall}, {Hewitson}, {Hild}, {Hirose}, {Hoak}, {Hodge}, {Holt},
  {Hosken}, {Hough}, {Howell}, {Hoyland}, {Huet}, {Hughey}, {Husa}, {Huttner},
  {Huynh-Dinh}, {Ingram}, {Inta}, {Isogai}, {Ivanov}, {Jaranowski}, {Johnson},
  {Jones}, {Jones}, {Jones}, {Ju}, {Kalmus}, {Kalogera}, {Kandhasamy},
  {Kanner}, {Katsavounidis}, {Kawabe}, {Kawamura}, {Kawazoe}, {Kells},
  {Keppel}, {Khalaidovski}, {Khalili}, {Khazanov}, {Kim}, {Kim}, {King},
  {Kinzel}, {Kissel}, {Klimenko}, {Kondrashov}, {Kopparapu}, {Koranda},
  {Kowalska}, {Kozak}, {Krause}, {Kringel}, {Krishnamurthy}, {Krishnan},
  {Kr{\'o}lak}, {Kuehn}, {Kullman}, {Kumar}, {Kwee}, {Landry}, {Lang}, {Lantz},
  {Lastzka}, {Lazzarini}, {Leaci}, {Leong}, {Leonor}, {Leroy}, {Letendre},
  {Li}, {Li}, {Lin}, {Lindquist}, {Lockerbie}, {Lodhia}, {Lorenzini},
  {Loriette}, {Lormand}, {Losurdo}, {Lu}, {Luan}, {Lubi{\'n}ski}, {Lucianetti},
  {L{\"u}ck}, {Lundgren}, {Machenschalk}, {MacInnis}, {Mackowski},
  {Mageswaran}, {Mailand}, {Majorana}, {Mak}, {Man}, {Mandel}, {Mandic},
  {Mantovani}, {Marchesoni}, {Marion}, {M{\'a}rka}, {M{\'a}rka}, {Maros},
  {Marque}, {Martelli}, {Martin}, {Martin}, {Marx}, {Mason}, {Masserot},
  {Matichard}, {Matone}, {Matzner}, {Mavalvala}, {McCarthy}, {McClelland},
  {McGuire}, {McIntyre}, {McIvor}, {McKechan}, {Meadors}, {Mehmet}, {Meier},
  {Melatos}, {Melissinos}, {Mendell}, {Men{\'e}ndez}, {Mercer}, {Merill},
  {Meshkov}, {Messenger}, {Meyer}, {Miao}, {Michel}, {Milano}, {Miller},
  {Minenkov}, {Mino}, {Mitra}, {Mitrofanov}, {Mitselmakher}, {Mittleman},
  {Moe}, {Mohan}, {Mohanty}, {Mohapatra}, {Moraru}, {Moreau}, {Moreno},
  {Morgado}, {Morgia}, {Morioka}, {Mors}, {Mosca}, {Moscatelli}, {Mossavi},
  {Mours}, {MowLowry}, {Mueller}, {Mukherjee}, {Mullavey},
  {M{\"u}ller-Ebhardt}, {Munch}, {Murray}, {Nash}, {Nawrodt}, {Nelson}, {Neri},
  {Newton}, {Nishizawa}, {Nocera}, {Nolting}, {Ochsner}, {O'Dell}, {Ogin},
  {Oldenburg}, {O'Reilly}, {O'Shaughnessy}, {Osthelder}, {Ottaway}, {Ottens},
  {Overmier}, {Owen}, {Page}, {Pagliaroli}, {Palladino}, {Palomba}, {Pan},
  {Pankow}, {Paoletti}, {Papa}, {Pardi}, {Pareja}, {Parisi}, {Pasqualetti},
  {Passaquieti}, {Passuello}, {Patel}, {Pedraza}, {Pekowsky}, {Penn},
  {Peralta}, {Perreca}, {Persichetti}, {Pichot}, {Pickenpack}, {Piergiovanni},
  {Pietka}, {Pinard}, {Pinto}, {Pitkin}, {Pletsch}, {Plissi}, {Poggiani},
  {Postiglione}, {Prato}, {Predoi}, {Price}, {Prijatelj}, {Principe},
  {Privitera}, {Prix}, {Prodi}, {Prokhorov}, {Puncken}, {Punturo}, {Puppo},
  {Quetschke}, {Raab}, {Rabaste}, {Rabeling}, {Radke}, {Radkins}, {Raffai},
  {Rakhmanov}, {Rankins}, {Rapagnani}, {Raymond}, {Re}, {Reed}, {Reed},
  {Regimbau}, {Reid}, {Reitze}, {Ricci}, {Riesen}, {Riles}, {Roberts},
  {Robertson}, {Robinet}, {Robinson}, {Robinson}, {Rocchi}, {Roddy},
  {R{\"o}ver}, {Rogstad}, {Rolland}, {Rollins}, {Romano}, {Romano}, {Romie},
  {Rosi{\'n}ska}, {Rowan}, {R{\"u}diger}, {Ruggi}, {Ryan}, {Sakata}, {Sakosky},
  {Salemi}, {Sammut}, {Sancho de la Jordana}, {Sandberg}, {Sannibale},
  {Santamar{\'\i}a}, {Santostasi}, {Saraf}, {Sassolas}, {Sathyaprakash},
  {Sato}, {Satterthwaite}, {Saulson}, {Savage}, {Schilling}, {Schnabel},
  {Schofield}, {Schulz}, {Schutz}, {Schwinberg}, {Scott}, {Scott}, {Searle},
  {Seifert}, {Sellers}, {Sengupta}, {Sentenac}, {Sergeev}, {Shaddock},
  {Shapiro}, {Shawhan}, {Shoemaker}, {Sibley}, {Siemens}, {Sigg}, {Singer},
  {Sintes}, {Skelton}, {Slagmolen}, {Slutsky}, {Smith}, {Smith}, {Smith},
  {Somiya}, {Sorazu}, {Speirits}, {Stein}, {Stein}, {Steinlechner},
  {Steplewski}, {Stochino}, {Stone}, {Strain}, {Strigin}, {Stroeer}, {Sturani},
  {Stuver}, {Summerscales}, {Sung}, {Susmithan}, {Sutton}, {Swinkels},
  {Talukder}, {Tanner}, {Tarabrin}, {Taylor}, {Taylor}, {Thomas}, {Thorne},
  {Thorne}, {Thrane}, {Th{\"u}ring}, {Titsler}, {Tokmakov}, {Toncelli},
  {Tonelli}, {Torres}, {Torrie}, {Tournefier}, {Travasso}, {Traylor}, {Trias},
  {Trummer}, {Tseng}, {Ugolini}, {Urbanek}, {Vahlbruch}, {Vaishnav}, {Vajente},
  {Vallisneri}, {van den Brand}, {Van Den Broeck}, {van der Putten}, {van der
  Sluys}, {van Veggel}, {Vass}, {Vaulin}, {Vavoulidis}, {Vecchio}, {Vedovato},
  {Veitch}, {Veitch}, {Veltkamp}, {Verkindt}, {Vetrano}, {Vicer{\'e}},
  {Villar}, {Vinet}, {Vocca}, {Vorvick}, {Vyachanin}, {Waldman}, {Wallace},
  {Wanner}, {Ward}, {Was}, {Wei}, {Weinert}, {Weinstein}, {Weiss}, {Wen},
  {Wen}, {Wessels}, {West}, {Westphal}, {Wette}, {Whelan}, {Whitcomb}, {White},
  {Whiting}, {Wilkinson}, {Willems}, {Williams}, {Willke}, {Winkelmann},
  {Winkler}, {Wipf}, {Wiseman}, {Woan}, {Wooley}, {Worden}, {Yakushin},
  {Yamamoto}, {Yamamoto}, {Yeaton-Massey}, {Yoshida}, {Yu}, {Yvert}, {Zanolin},
  {Zhang}, {Zhang}, {Zhao}, {Zotov}, {Zucker}, {Zweizig}, {LIGO Scientific
  Collaboration}, \& {Virgo Collaboration}}]{2010CQGra..27q3001A}
{Abadie}, J., {Abbott}, B.~P., {Abbott}, R., {et~al.} 2010, Classical and
  Quantum Gravity, 27, 173001

\bibitem[{{Abbott} {et~al.}(2016){Abbott}, {Abbott}, {Abbott}, {Abernathy},
  {Acernese}, {Ackley}, {Adams}, {Adams}, {Addesso}, {Adhikari}, {Adya},
  {Affeldt}, {Agathos}, {Agatsuma}, {Aggarwal}, {Aguiar}, {Aiello}, {Ain},
  {Ajith}, {Allen}, {Allocca}, {Altin}, {Anderson}, {Anderson}, {Arai},
  {Araya}, {Arceneaux}, {Areeda}, {Arnaud}, {Arun}, {Ascenzi}, {Ashton}, {Ast},
  {Aston}, {Astone}, {Aufmuth}, {Aulbert}, {Babak}, {Bacon}, {Bader}, {Baker},
  {Baldaccini}, {Ballardin}, {Ballmer}, {Barayoga}, {Barclay}, {Barish},
  {Barker}, {Barone}, {Barr}, {Barsotti}, {Barsuglia}, {Barta}, {Bartlett},
  {Bartos}, {Bassiri}, {Basti}, {Batch}, {Baune}, {Bavigadda}, {Bazzan},
  {Behnke}, {Bejger}, {Belczynski}, {Bell}, {Bell}, {Berger}, {Bergman},
  {Bergmann}, {Berry}, {Bersanetti}, {Bertolini}, {Betzwieser}, {Bhagwat},
  {Bhandare}, {Bilenko}, {Billingsley}, {Birch}, {Birney}, {Biscans}, {Bisht},
  {Bitossi}, {Biwer}, {Bizouard}, {Blackburn}, {Blair}, {Blair}, {Blair},
  {Bloemen}, {Bock}, {Bodiya}, {Boer}, {Bogaert}, {Bogan}, {Bohe}, {Bojtos},
  {Bond}, {Bondu}, {Bonnand}, {Boom}, {Bork}, {Boschi}, {Bose}, {Bouffanais},
  {Bozzi}, {Bradaschia}, {Brady}, {Braginsky}, {Branchesi}, {Brau}, {Briant},
  {Brillet}, {Brinkmann}, {Brisson}, {Brockill}, {Brooks}, {Brown}, {Brown},
  {Brown}, {Buchanan}, {Buikema}, {Bulik}, {Bulten}, {Buonanno}, {Buskulic},
  {Buy}, {Byer}, {Cadonati}, {Cagnoli}, {Cahillane}, {Calder{\'o}n Bustillo},
  {Callister}, {Calloni}, {Camp}, {Cannon}, {Cao}, {Capano}, {Capocasa},
  {Carbognani}, {Caride}, {Casanueva Diaz}, {Casentini}, {Caudill},
  {Cavagli{\`a}}, {Cavalier}, {Cavalieri}, {Cella}, {Cepeda}, {Cerboni
  Baiardi}, {Cerretani}, {Cesarini}, {Chakraborty}, {Chalermsongsak},
  {Chamberlin}, {Chan}, {Chao}, {Charlton}, {Chassande-Mottin}, {Chen}, {Chen},
  {Cheng}, {Chincarini}, {Chiummo}, {Cho}, {Cho}, {Chow}, {Christensen}, {Chu},
  {Chua}, {Chung}, {Ciani}, {Clara}, {Clark}, {Cleva}, {Coccia}, {Cohadon},
  {Colla}, {Collette}, {Cominsky}, {Constancio}, {Conte}, {Conti}, {Cook},
  {Corbitt}, {Cornish}, {Corsi}, {Cortese}, {Costa}, {Coughlin}, {Coughlin},
  {Coulon}, {Countryman}, {Couvares}, {Cowan}, {Coward}, {Cowart}, {Coyne},
  {Coyne}, {Craig}, {Creighton}, {Cripe}, {Crowder}, {Cumming}, {Cunningham},
  {Cuoco}, {Dal Canton}, {Danilishin}, {D'Antonio}, {Danzmann}, {Darman},
  {Dattilo}, {Dave}, {Daveloza}, {Davier}, {Davies}, {Daw}, {Day}, {DeBra},
  {Debreczeni}, {Degallaix}, {De Laurentis}, {Del{\'e}glise}, {Del Pozzo},
  {Denker}, {Dent}, {Dereli}, {Dergachev}, {DeRosa}, {DeRosa}, {DeSalvo},
  {Dhurandhar}, {D{\'\i}az}, {Di Fiore}, {Di Giovanni}, {Di Lieto}, {Di Pace},
  {Di Palma}, {Di Virgilio}, {Dojcinoski}, {Dolique}, {Donovan}, {Dooley},
  {Doravari}, {Douglas}, {Downes}, {Drago}, {Drever}, {Driggers}, {Du},
  {Ducrot}, {Dwyer}, {Edo}, {Edwards}, {Effler}, {Eggenstein}, {Ehrens},
  {Eichholz}, {Eikenberry}, {Engels}, {Essick}, {Etzel}, {Evans}, {Evans},
  {Everett}, {Factourovich}, {Fafone}, {Fair}, {Fairhurst}, {Fan}, {Fang},
  {Farinon}, {Farr}, {Farr}, {Favata}, {Fays}, {Fehrmann}, {Fejer}, {Ferrante},
  {Ferreira}, {Ferrini}, {Fidecaro}, {Fiori}, {Fiorucci}, {Fisher}, {Flaminio},
  {Fletcher}, {Fournier}, {Franco}, {Frasca}, {Frasconi}, {Frei}, {Freise},
  {Frey}, {Frey}, {Fricke}, {Fritschel}, {Frolov}, {Fulda}, {Fyffe}, {Gabbard},
  {Gair}, {Gammaitoni}, {Gaonkar}, {Garufi}, {Gatto}, {Gaur}, {Gehrels},
  {Gemme}, {Gendre}, {Genin}, {Gennai}, {George}, {Gergely}, {Germain},
  {Ghosh}, {Ghosh}, {Giaime}, {Giardina}, {Giazotto}, {Gill}, {Glaefke},
  {Goetz}, {Goetz}, {Gondan}, {Gonz{\'a}lez}, {Gonzalez Castro}, {Gopakumar},
  {Gordon}, {Gorodetsky}, {Gossan}, {Gosselin}, {Gouaty}, {Graef}, {Graff},
  {Granata}, {Grant}, {Gras}, {Gray}, {Greco}, {Green}, {Groot}, {Grote},
  {Grunewald}, {Guidi}, {Guo}, {Gupta}, {Gupta}, {Gushwa}, {Gustafson},
  {Gustafson}, {Hacker}, {Hall}, {Hall}, {Hammond}, {Haney}, {Hanke}, {Hanks},
  {Hanna}, {Hannam}, {Hanson}, {Hardwick}, {Harms}, {Harry}, {Harry}, {Hart},
  {Hartman}, {Haster}, {Haughian}, {Heidmann}, {Heintze}, {Heitmann}, {Hello},
  {Hemming}, {Hendry}, {Heng}, {Hennig}, {Heptonstall}, {Heurs}, {Hild},
  {Hoak}, {Hodge}, {Hofman}, {Hollitt}, {Holt}, {Holz}, {Hopkins}, {Hosken},
  {Hough}, {Houston}, {Howell}, {Hu}, {Huang}, {Huerta}, {Huet}, {Hughey},
  {Husa}, {Huttner}, {Huynh-Dinh}, {Idrisy}, {Indik}, {Ingram}, {Inta}, {Isa},
  {Isac}, {Isi}, {Islas}, {Isogai}, {Iyer}, {Izumi}, {Jacqmin}, {Jang}, {Jani},
  {Jaranowski}, {Jawahar}, {Jim{\'e}nez-Forteza}, {Johnson}, {Jones}, {Jones},
  {Jonker}, {Ju}, {K}, {Kalaghatgi}, {Kalogera}, {Kandhasamy}, {Kang},
  {Kanner}, {Karki}, {Kasprzack}, {Katsavounidis}, {Katzman}, {Kaufer}, {Kaur},
  {Kawabe}, {Kawazoe}, {K{\'e}f{\'e}lian}, {Kehl}, {Keitel}, {Kelley}, {Kells},
  {Kennedy}, {Key}, {Khalaidovski}, {Khalili}, {Khan}, {Khan}, {Khan},
  {Khazanov}, {Kijbunchoo}, {Kim}, {Kim}, {Kim}, {Kim}, {Kim}, {Kim}, {King},
  {King}, {Kinzel}, {Kissel}, {Kleybolte}, {Klimenko}, {Koehlenbeck},
  {Kokeyama}, {Koley}, {Kondrashov}, {Kontos}, {Korobko}, {Korth}, {Kowalska},
  {Kozak}, {Kringel}, {Krishnan}, {Kr{\'o}lak}, {Krueger}, {Kuehn}, {Kumar},
  {Kuo}, {Kutynia}, {Lackey}, {Landry}, {Lange}, {Lantz}, {Lasky}, {Lazzarini},
  {Lazzaro}, {Leaci}, {Leavey}, {Lebigot}, {Lee}, {Lee}, {Lee}, {Lee}, {Lenon},
  {Leonardi}, {Leong}, {Leroy}, {Letendre}, {Levin}, {Levine}, {Li}, {Libson},
  {Littenberg}, {Lockerbie}, {Logue}, {Lombardi}, {Lord}, {Lorenzini},
  {Loriette}, {Lormand}, {Losurdo}, {Lough}, {L{\"u}ck}, {Lundgren}, {Luo},
  {Lynch}, {Ma}, {MacDonald}, {Machenschalk}, {MacInnis}, {Macleod},
  {Maga{\~n}a-Sandoval}, {Magee}, {Mageswaran}, {Majorana}, {Maksimovic},
  {Malvezzi}, {Man}, {Mandel}, {Mandic}, {Mangano}, {Mansell}, {Manske},
  {Mantovani}, {Marchesoni}, {Marion}, {M{\'a}rka}, {M{\'a}rka}, {Markosyan},
  {Maros}, {Martelli}, {Martellini}, {Martin}, {Martin}, {Martynov}, {Marx},
  {Mason}, {Masserot}, {Massinger}, {Masso-Reid}, {Matichard}, {Matone},
  {Mavalvala}, {Mazumder}, {Mazzolo}, {McCarthy}, {McClelland}, {McCormick},
  {McGuire}, {McIntyre}, {McIver}, {McManus}, {McWilliams}, {Meacher},
  {Meadors}, {Meidam}, {Melatos}, {Mendell}, {Mendoza-Gandara}, {Mercer},
  {Merilh}, {Merzougui}, {Meshkov}, {Messenger}, {Messick}, {Meyers},
  {Mezzani}, {Miao}, {Michel}, {Middleton}, {Mikhailov}, {Milano}, {Miller},
  {Millhouse}, {Minenkov}, {Ming}, {Mirshekari}, {Mishra}, {Mitra},
  {Mitrofanov}, {Mitselmakher}, {Mittleman}, {Moggi}, {Mohan}, {Mohapatra},
  {Montani}, {Moore}, {Moore}, {Moraru}, {Moreno}, {Morriss}, {Mossavi},
  {Mours}, {Mow-Lowry}, {Mueller}, {Mueller}, {Muir}, {Mukherjee}, {Mukherjee},
  {Mukherjee}, {Mukund}, {Mullavey}, {Munch}, {Murphy}, {Murray}, {Mytidis},
  {Nardecchia}, {Naticchioni}, {Nayak}, {Necula}, {Nedkova}, {Nelemans},
  {Neri}, {Neunzert}, {Newton}, {Nguyen}, {Nielsen}, {Nissanke}, {Nitz},
  {Nocera}, {Nolting}, {Normandin}, {Nuttall}, {Oberling}, {Ochsner}, {O'Dell},
  {Oelker}, {Ogin}, {Oh}, {Oh}, {Ohme}, {Oliver}, {Oppermann}, {Oram},
  {O'Reilly}, {O'Shaughnessy}, {Ottaway}, {Ottens}, {Overmier}, {Owen}, {Pai},
  {Pai}, {Palamos}, {Palashov}, {Palomba}, {Pal-Singh}, {Pan}, {Pankow},
  {Pannarale}, {Pant}, {Paoletti}, {Paoli}, {Papa}, {Paris}, {Parker},
  {Pascucci}, {Pasqualetti}, {Passaquieti}, {Passuello}, {Patricelli},
  {Patrick}, {Pearlstone}, {Pedraza}, {Pedurand}, {Pekowsky}, {Pele}, {Penn},
  {Perreca}, {Phelps}, {Piccinni}, {Pichot}, {Piergiovanni}, {Pierro},
  {Pillant}, {Pinard}, {Pinto}, {Pitkin}, {Poggiani}, {Popolizio}, {Post},
  {Powell}, {Prasad}, {Predoi}, {Premachandra}, {Prestegard}, {Price},
  {Prijatelj}, {Principe}, {Privitera}, {Prix}, {Prodi}, {Prokhorov},
  {Puncken}, {Punturo}, {Puppo}, {P{\"u}rrer}, {Qi}, {Qin}, {Quetschke},
  {Quintero}, {Quitzow-James}, {Raab}, {Rabeling}, {Radkins}, {Raffai}, {Raja},
  {Rakhmanov}, {Rapagnani}, {Raymond}, {Razzano}, {Re}, {Read}, {Reed},
  {Regimbau}, {Rei}, {Reid}, {Reitze}, {Rew}, {Reyes}, {Ricci}, {Riles},
  {Robertson}, {Robie}, {Robinet}, {Rocchi}, {Rolland}, {Rollins}, {Roma},
  {Romano}, {Romano}, {Romanov}, {Romie}, {Rosi{\'n}ska}, {Rowan},
  {R{\"u}diger}, {Ruggi}, {Ryan}, {Sachdev}, {Sadecki}, {Sadeghian}, {Salconi},
  {Saleem}, {Salemi}, {Samajdar}, {Sammut}, {Sanchez}, {Sandberg}, {Sandeen},
  {Sanders}, {Sassolas}, {Sathyaprakash}, {Saulson}, {Sauter}, {Savage},
  {Sawadsky}, {Schale}, {Schilling}, {Schmidt}, {Schmidt}, {Schnabel},
  {Schofield}, {Sch{\"o}nbeck}, {Schreiber}, {Schuette}, {Schutz}, {Scott},
  {Scott}, {Sellers}, {Sentenac}, {Sequino}, {Sergeev}, {Serna}, {Setyawati},
  {Sevigny}, {Shaddock}, {Shah}, {Shahriar}, {Shaltev}, {Shao}, {Shapiro},
  {Shawhan}, {Sheperd}, {Shoemaker}, {Shoemaker}, {Siellez}, {Siemens}, {Sigg},
  {Silva}, {Simakov}, {Singer}, {Singer}, {Singh}, {Singh}, {Singhal},
  {Sintes}, {Slagmolen}, {Smith}, {Smith}, {Smith}, {Son}, {Sorazu},
  {Sorrentino}, {Souradeep}, {Srivastava}, {Staley}, {Steinke}, {Steinlechner},
  {Steinlechner}, {Steinmeyer}, {Stephens}, {Stevenson}, {Stone}, {Strain},
  {Straniero}, {Stratta}, {Strauss}, {Strigin}, {Sturani}, {Stuver},
  {Summerscales}, {Sun}, {Sutton}, {Swinkels}, {Szczepa{\'n}czyk}, {Tacca},
  {Talukder}, {Tanner}, {T{\'a}pai}, {Tarabrin}, {Taracchini}, {Taylor},
  {Theeg}, {Thirugnanasambandam}, {Thomas}, {Thomas}, {Thomas}, {Thorne},
  {Thorne}, {Thrane}, {Tiwari}, {Tiwari}, {Tokmakov}, {Tomlinson}, {Tonelli},
  {Torres}, {Torrie}, {T{\"o}yr{\"a}}, {Travasso}, {Traylor}, {Trifir{\`o}},
  {Tringali}, {Trozzo}, {Tse}, {Turconi}, {Tuyenbayev}, {Ugolini},
  {Unnikrishnan}, {Urban}, {Usman}, {Vahlbruch}, {Vajente}, {Valdes}, {van
  Bakel}, {van Beuzekom}, {van den Brand}, {van den Broeck}, {Vander-Hyde},
  {van der Schaaf}, {van Heijningen}, {van Veggel}, {Vardaro}, {Vass},
  {Vas{\'u}th}, {Vaulin}, {Vecchio}, {Vedovato}, {Veitch}, {Veitch},
  {Venkateswara}, {Verkindt}, {Vetrano}, {Vicer{\'e}}, {Vinciguerra}, {Vine},
  {Vinet}, {Vitale}, {Vo}, {Vocca}, {Vorvick}, {Voss}, {Vousden}, {Vyatchanin},
  {Wade}, {Wade}, {Wade}, {Walker}, {Wallace}, {Walsh}, {Wang}, {Wang}, {Wang},
  {Wang}, {Wang}, {Ward}, {Warner}, {Was}, {Weaver}, {Wei}, {Weinert},
  {Weinstein}, {Weiss}, {Welborn}, {Wen}, {We{\ss}els}, {Westphal}, {Wette},
  {Whelan}, {White}, {Whiting}, {Williams}, {Williamson}, {Willis}, {Willke},
  {Wimmer}, {Winkler}, {Wipf}, {Wittel}, {Woan}, {Worden}, {Wright}, {Wu},
  {Yablon}, {Yam}, {Yamamoto}, {Yancey}, {Yap}, {Yu}, {Yvert}, {Zadro{\.z}ny},
  {Zangrando}, {Zanolin}, {Zendri}, {Zevin}, {Zhang}, {Zhang}, {Zhang},
  {Zhang}, {Zhao}, {Zhou}, {Zhou}, {Zhu}, {Zucker}, {Zuraw}, {and}, {Zweizig},
  {LIGO Scientific Collaboration}, \& {Virgo
  Collaboration}}]{2016ApJ...818L..22A}
{Abbott}, B.~P., {Abbott}, R., {Abbott}, T.~D., {et~al.} 2016, \apjl, 818, L22

\bibitem[{Abbott {et~al.}(2021)Abbott, Abbott, Abraham, Acernese, Ackley,
  Adams, Adams, Adhikari, Adya, Affeldt, Agarwal, Agathos, Agatsuma, Aggarwal,
  Aguiar, Aiello, Ain, Akutsu, Aleman, Allen, Allocca, Altin, Amato, Anand,
  Ananyeva, Anderson, Anderson, Ando, Angelova, Ansoldi, Antelis, Antier,
  Appert, Arai, Arai, Arai, Araki, Araya, Araya, Areeda, Ar\`ene, Aritomi,
  Arnaud, Aronson, Asada, Asali, Ashton, Aso, Aston, Astone, Aubin, Aufmuth,
  AultONeal, Austin, Babak, Badaracco, Bader, Bae, Bae, Baer, Bagnasco, Bai,
  Baiotti, Baird, Bajpai, Ball, Ballardin, Ballmer, Bals, Balsamo, Baltus,
  Banagiri, Bankar, Bankar, Barayoga, Barbieri, Barish, Barker, Barneo, Barnum,
  Barone, Barr, Barsotti, Barsuglia, Barta, Bartlett, Barton, Bartos, Bassiri,
  Basti, Bawaj, Bayley, Baylor, Bazzan, B\'ecsy, Bedakihale, Bejger, Belahcene,
  Benedetto, Beniwal, Benjamin, Bennett, Bentley, BenYaala, Bergamin, Berger,
  Bernuzzi, Bersanetti, Bertolini, Betzwieser, Bhandare, Bhandari,
  Bhattacharjee, Bhaumik, Bidler, Bilenko, Billingsley, Birney, Birnholtz,
  Biscans, Bischi, Biscoveanu, Bisht, Biswas, Bitossi, Bizouard, Blackburn,
  Blackman, Blair, Blair, Blair, Bobba, Bode, Boer, Bogaert, Boldrini, Bondu,
  Bonilla, Bonnand, Booker, Boom, Bork, Boschi, Bose, Bose, Bossilkov, Boudart,
  Bouffanais, Bozzi, Bradaschia, Brady, Bramley, Branch, Branchesi, Brau,
  Breschi, Briant, Briggs, Brillet, Brinkmann, Brockill, Brooks, Brooks, Brown,
  Brunett, Bruno, Bruntz, Bryant, Buikema, Bulik, Bulten, Buonanno, Buscicchio,
  Buskulic, Byer, Cadonati, Caesar, Cagnoli, Cahillane, Cain, Bustillo,
  Callaghan, Callister, Calloni, Camp, Canepa, Cannavacciuolo, Cannon, Cao,
  Cao, Cao, Capocasa, Capote, Carapella, Carbognani, Carlin, Carney,
  Carpinelli, Carullo, Carver, Diaz, Casentini, Castaldi, Caudill, Cavagli\`a,
  Cavalier, Cavalieri, Cella, Cerd\'a-Dur\'an, Cesarini, Chaibi, Chakravarti,
  Champion, Chan, Chan, Chan, Chan, Chandra, Chanial, Chao, Charlton, Chase,
  Chassande-Mottin, Chatterjee, Chaturvedi, Chen, Chen, Chen, Chen, Chen, Chen,
  Chen, Chen, Chen, Cheng, Cheong, Cheung, Chia, Chiadini, Chiang, Chierici,
  Chincarini, Chiofalo, Chiummo, Cho, Cho, Choate, Choudhary, Choudhary,
  Christensen, Chu, Chu, Chu, Chua, Chung, Ciani, Ciecielag,
  Cie\ifmmode~\acute{s}\else \'{s}\fi{}lar, Cifaldi, Ciobanu, Ciolfi, Cipriano,
  Cirone, Clara, Clark, Clark, Clarke, Clearwater, Clesse, Cleva, Coccia,
  Cohadon, Cohen, Cohen, Colleoni, Collette, Colpi, Compton, Constancio, Conti,
  Cooper, Corban, Corbitt, Cordero-Carri\'on, Corezzi, Corley, Cornish, Corre,
  Corsi, Cortese, Costa, Cotesta, Coughlin, Coughlin, Coulon, Countryman,
  Cousins, Couvares, Covas, Coward, Cowart, Coyne, Coyne, Creighton, Creighton,
  Criswell, Croquette, Crowder, Cudell, Cullen, Cumming, Cummings, Cuoco,
  Cury\l{}o, Canton, D\'alya, Dana, DaneshgaranBajastani, D'Angelo, Danilishin,
  D'Antonio, Danzmann, Darsow-Fromm, Dasgupta, Datrier, Dattilo, Dave, Davier,
  Davies, Davis, Daw, Dean, Deenadayalan, Degallaix, De~Laurentis, Del\'eglise,
  Del~Favero, De~Lillo, De~Lillo, Del~Pozzo, DeMarchi, De~Matteis, D'Emilio,
  Demos, Dent, Depasse, De~Pietri, De~Rosa, De~Rossi, DeSalvo, De~Simone,
  Dhurandhar, D\'{\i}az, Diaz-Ortiz, Didio, Dietrich, Di~Fiore, Di~Fronzo,
  Di~Giorgio, Di~Giovanni, Di~Girolamo, Di~Lieto, Ding, Di~Pace, Di~Palma,
  Di~Renzo, Divakarla, Dmitriev, Doctor, D'Onofrio, Donovan, Dooley, Doravari,
  Dorrington, Drago, Driggers, Drori, Du, Ducoin, Dupej, Durante, D'Urso,
  Duverne, Dvorkin, Dwyer, Easter, Ebersold, Eddolls, Edelman, Edo, Edy,
  Effler, Eguchi, Eichholz, Eikenberry, Eisenmann, Eisenstein, Ejlli, Enomoto,
  Errico, Essick, Estell\'es, Estevez, Etienne, Etzel, Evans, Evans, Ewing,
  Fafone, Fair, Fairhurst, Fan, Farah, Farinon, Farr, Farr, Farrow,
  Fauchon-Jones, Favata, Fays, Fazio, Feicht, Fejer, Feng, Fenyvesi, Ferguson,
  Fernandez-Galiana, Ferrante, Ferreira, Fidecaro, Figura, Fiori, Fishbach,
  Fisher, Fishner, Fittipaldi, Fiumara, Flaminio, Floden, Flynn, Fong, Font,
  Fornal, Forsyth, Franke, Frasca, Frasconi, Frederick, Frei, Freise, Frey,
  Fritschel, Frolov, Fronz\'e, Fujii, Fujikawa, Fukunaga, Fukushima, Fulda,
  Fyffe, Gabbard, Gadre, Gaebel, Gair, Gais, Galaudage, Gamba, Ganapathy,
  Ganguly, Gao, Gaonkar, Garaventa, Garc\'{\i}a-N\'u\~nez,
  Garc\'{\i}a-Quir\'os, Garufi, Gateley, Gaudio, Gayathri, Ge, Gemme, Gennai,
  George, Gergely, Gewecke, Ghonge, Ghosh, Ghosh, Ghosh, Ghosh, Ghosh,
  Giacomazzo, Giacoppo, Giaime, Giardina, Gibson, Gier, Giesler, Giri, Gissi,
  Glanzer, Gleckl, Godwin, Goetz, Goetz, Gohlke, Goncharov, Gonz\'alez,
  Gopakumar, Gosselin, Gouaty, Grace, Grado, Granata, Granata, Grant, Gras,
  Grassia, Gray, Gray, Greco, Green, Green, Gretarsson, Gretarsson, Griffith,
  Griffiths, Griggs, Grignani, Grimaldi, Grimes, Grimm, Grote, Grunewald,
  Gruning, Guerrero, Guidi, Guimaraes, Guix\'e, Gulati, Guo, Guo, Gupta, Gupta,
  Gupta, Gustafson, Gustafson, Guzman, Ha, Haegel, Hagiwara, Haino, Halim,
  Hall, Hamilton, Hammond, Han, Haney, Hanks, Hanna, Hannam, Hannuksela,
  Hansen, Hansen, Hanson, Harder, Hardwick, Haris, Harms, Harry, Harry,
  Hartwig, Hasegawa, Haskell, Hasskew, Haster, Hattori, Haughian, Hayakawa,
  Hayama, Hayes, Healy, Heidmann, Heintze, Heinze, Heinzel, Heitmann, Hellman,
  Hello, Helmling-Cornell, Hemming, Hendry, Heng, Hennes, Hennig, Hennig,
  Vivanco, Heurs, Hild, Hill, Himemoto, Hines, Hiranuma, Hirata, Hirose,
  Hochheim, Hofman, Hohmann, Holgado, Holland, Hollows, Holmes, Holt, Holz,
  Hong, Hopkins, Hough, Howell, Hoy, Hoyland, Hreibi, Hsieh, Hsu, Huang, Huang,
  Huang, Huang, Huang, Huang, H\"ubner, Huddart, Huerta, Hughey, Hui, Hui,
  Husa, Huttner, Huxford, Huynh-Dinh, Ide, Idzkowski, Iess, Ikenoue, Imam,
  Inayoshi, Inchauspe, Ingram, Inoue, Intini, Ioka, Isi, Isleif, Ito, Itoh,
  Iyer, Izumi, JaberianHamedan, Jacqmin, Jadhav, Jadhav, James, Jan, Jani,
  Janssens, Janthalur, Jaranowski, Jariwala, Jaume, Jenkins, Jeon, Jeunon, Jia,
  Jiang, Jin, Johns, Jones, Jones, Jones, Jones, Jones, Jonker, Ju, Jung, Jung,
  Junker, Kaihotsu, Kajita, Kakizaki, Kalaghatgi, Kalogera, Kamai, Kamiizumi,
  Kanda, Kandhasamy, Kang, Kanner, Kao, Kapadia, Kapasi, Karathanasis, Karki,
  Kashyap, Kasprzack, Kastaun, Katsanevas, Katsavounidis, Katzman, Kaur,
  Kawabe, Kawaguchi, Kawai, Kawasaki, K\'ef\'elian, Keitel, Key, Khadka,
  Khalili, Khan, Khan, Khazanov, Khetan, Khursheed, Kijbunchoo, Kim, Kim, Kim,
  Kim, Kim, Kim, Kimball, Kimura, King, Kinley-Hanlon, Kirchhoff, Kissel, Kita,
  Kitazawa, Kleybolte, Klimenko, Knee, Knowles, Knyazev, Koch, Koekoek, Kojima,
  Kokeyama, Koley, Kolitsidou, Kolstein, Komori, Kondrashov, Kong, Kontos,
  Koper, Korobko, Kotake, Kovalam, Kozak, Kozakai, Kozu, Kringel, Krishnendu,
  Kr\'olak, Kuehn, Kuei, Kumar, Kumar, Kumar, Kumar, Kume, Kuns, Kuo, Kuo,
  Kuromiya, Kuroyanagi, Kusayanagi, Kwak, Kwang, Laghi, Lalande, Lam, Lamberts,
  Landry, Lane, Lang, Lange, Lantz, La~Rosa, Lartaux-Vollard, Lasky, Laxen,
  Lazzarini, Lazzaro, Leaci, Leavey, Lecoeuche, Lee, Lee, Lee, Lee, Lee, Lee,
  Lehmann, Lema\^{\i}tre, Leon, Leonardi, Leroy, Letendre, Levin, Leviton, Li,
  Li, Li, Li, Li, Li, Lin, Lin, Lin, Lin, Lin, Linde, Linker, Linley,
  Littenberg, Liu, Liu, Liu, Liu, Llorens-Monteagudo, Lo, Lockwood, Lollie,
  London, Longo, Lopez, Lorenzini, Loriette, Lormand, Losurdo, Lough, Lousto,
  Lovelace, L\"uck, Lumaca, Lundgren, Luo, Macas, MacInnis, Macleod, MacMillan,
  Macquet, Hernandez, Maga\~na Sandoval, Magazz\`u, Magee, Maggiore, Majorana,
  Maksimovic, Maliakal, Malik, Man, Mandic, Mangano, Mango, Mansell, Manske,
  Mantovani, Mapelli, Marchesoni, Marchio, Marion, Mark, M\'arka, M\'arka,
  Markakis, Markosyan, Markowitz, Maros, Marquina, Marsat, Martelli, Martin,
  Martin, Martinez, Martinez, Martinovic, Martynov, Marx, Masalehdan, Mason,
  Massera, Masserot, Massinger, Masso-Reid, Mastrogiovanni, Matas,
  Mateu-Lucena, Matichard, Matiushechkina, Mavalvala, McCann, McCarthy,
  McClelland, McClincy, McCormick, McCuller, McGhee, McGuire, McIsaac, McIver,
  McManus, McRae, McWilliams, Meacher, Mehmet, Mehta, Melatos, Melchor,
  Mendell, Menendez-Vazquez, Menoni, Mercer, Mereni, Merfeld, Merilh, Merritt,
  Merzougui, Meshkov, Messenger, Messick, Meyers, Meylahn, Mhaske, Miani, Miao,
  Michaloliakos, Michel, Michimura, Middleton, Milano, Miller, Millhouse,
  Mills, Milotti, Milovich-Goff, Minazzoli, Minenkov, Mio, Mir, Mishkin,
  Mishra, Mishra, Mistry, Mitra, Mitrofanov, Mitselmakher, Mittleman, Miyakawa,
  Miyamoto, Miyazaki, Miyo, Miyoki, Mo, Mogushi, Mohapatra, Mohite, Molina,
  Molina-Ruiz, Mondin, Montani, Moore, Moraru, Morawski, More, Moreno, Moreno,
  Mori, Morisaki, Moriwaki, Mours, Mow-Lowry, Mozzon, Muciaccia, Mukherjee,
  Mukherjee, Mukherjee, Mukherjee, Mukund, Mullavey, Munch, Mu\~niz, Murray,
  Musenich, Nadji, Nagano, Nagano, Nagar, Nakamura, Nakano, Nakano, Nakashima,
  Nakayama, Nardecchia, Narikawa, Naticchioni, Nayak, Nayak, Negishi, Neil,
  Neilson, Nelemans, Nelson, Nery, Neunzert, Ng, Ng, Nguyen, Nguyen, Nguyen,
  Quynh, Ni, Nichols, Nishizawa, Nissanke, Nocera, Noh, Norman, North, Nozaki,
  Nuttall, Oberling, O'Brien, Obuchi, O'Dell, Ogaki, Oganesyan, Oh, Oh, Oh,
  Ohashi, Ohishi, Ohkawa, Ohme, Ohta, Okada, Okutani, Okutomi, Olivetto,
  Oohara, Ooi, Oram, O'Reilly, Ormiston, Ormsby, Ortega, O'Shaughnessy, O'Shea,
  Oshino, Ossokine, Osthelder, Otabe, Ottaway, Overmier, Pace, Pagano, Page,
  Pagliaroli, Pai, Pai, Palamos, Palashov, Palomba, Pan, Panda, Pang, Pang,
  Pankow, Pannarale, Pant, Paoletti, Paoli, Paolone, Parisi, Park, Parker,
  Pascucci, Pasqualetti, Passaquieti, Passuello, Patel, Patricelli, Payne,
  Pechsiri, Pedraza, Pegoraro, Pele, Arellano, Penn, Perego, Pereira, Pereira,
  Perez, P\'erigois, Perreca, Perri\`es, Petermann, Petterson, Pfeiffer, Pham,
  Phukon, Piccinni, Pichot, Piendibene, Piergiovanni, Pierini, Pierro, Pillant,
  Pilo, Pinard, Pinto, Piotrzkowski, Piotrzkowski, Pirello, Pitkin, Placidi,
  Plastino, Pluchar, Poggiani, Polini, Pong, Ponrathnam, Popolizio, Porter,
  Powell, Pracchia, Pradier, Prajapati, Prasai, Prasanna, Pratten, Prestegard,
  Principe, Prodi, Prokhorov, Prosposito, Prudenzi, Puecher, Punturo, Puosi,
  Puppo, P\"urrer, Qi, Quetschke, Quinonez, Quitzow-James, Raab, Raaijmakers,
  Radkins, Radulesco, Raffai, Rail, Raja, Rajan, Ramirez, Ramirez,
  Ramos-Buades, Rana, Rapagnani, Rapol, Ratto, Raymond, Raza, Razzano, Read,
  Rees, Regimbau, Rei, Reid, Reitze, Relton, Rettegno, Ricci, Richardson,
  Richardson, Richardson, Ricker, Riemenschneider, Riles, Rizzo, Robertson,
  Robie, Robinet, Rocchi, Rocha, Rodriguez, Rodriguez-Soto, Rolland, Rollins,
  Roma, Romanelli, Romano, Romano, Romel, Romero, Romero-Shaw, Romie, Rose,
  Rosi\ifmmode~\acute{n}\else \'{n}\fi{}ska, Rosofsky, Ross, Rowan, Rowlinson,
  Roy, Roy, Rozza, Ruggi, Ryan, Sachdev, Sadecki, Sadiq, Sago, Saito, Saito,
  Sakai, Sakai, Sakellariadou, Sakuno, Salafia, Salconi, Saleem, Salemi,
  Samajdar, Sanchez, Sanchez, Sanchez, Sanchis-Gual, Sanders, Sanuy, Saravanan,
  Sarin, Sassolas, Satari, Sato, Sato, Sauter, Savage, Savant, Sawada, Sawant,
  Sawant, Sayah, Schaetzl, Scheel, Scheuer, Schindler-Tyka, Schmidt, Schnabel,
  Schneewind, Schofield, Sch\"onbeck, Schulte, Schutz, Schwartz, Scott, Scott,
  Seglar-Arroyo, Seidel, Sekiguchi, Sekiguchi, Sellers, Sengupta, Sennett,
  Sentenac, Seo, Sequino, Sergeev, Setyawati, Shaffer, Shahriar, Shams, Shao,
  Sharifi, Sharma, Sharma, Shawhan, Shcheblanov, Shen, Shibagaki, Shikauchi,
  Shimizu, Shimoda, Shimode, Shink, Shinkai, Shishido, Shoda, Shoemaker,
  Shoemaker, Shukla, ShyamSundar, Sieniawska, Sigg, Singer, Singh, Singh,
  Singha, Sintes, Sipala, Skliris, Slagmolen, Slaven-Blair, Smetana, Smith,
  Smith, Somala, Somiya, Son, Soni, Soni, Sorazu, Sordini, Sorrentino,
  Sorrentino, Sotani, Soulard, Souradeep, Sowell, Spagnuolo, Spencer, Spera,
  Srivastava, Srivastava, Staats, Stachie, Steer, Steinlechner, Steinlechner,
  Stops, Stover, Strain, Strang, Stratta, Strunk, Sturani, Stuver, S\"udbeck,
  Sudhagar, Sudhir, Sugimoto, Suh, Summerscales, Sun, Sun, Sunil, Sur, Suresh,
  Sutton, Suzuki, Suzuki, Swinkels, Szczepa\ifmmode~\acute{n}\else
  \'{n}\fi{}czyk, Szewczyk, Tacca, Tagoshi, Tait, Takahashi, Takahashi,
  Takamori, Takano, Takeda, Takeda, Talbot, Tanaka, Tanaka, Tanaka, Tanaka,
  Tanaka, Tanasijczuk, Tanioka, Tanner, Tao, Tapia, Martin, Martin, Tasson,
  Telada, Tenorio, Terkowski, Test, Thirugnanasambandam, Thomas, Thomas,
  Thompson, Thondapu, Thorne, Thrane, Tiwari, Tiwari, Tiwari, Toland, Tolley,
  Tomaru, Tomigami, Tomura, Tonelli, Torres-Forn\'e, Torrie, e~Melo, T\"oyr\"a,
  Trapananti, Travasso, Traylor, Tringali, Tripathee, Troiano, Trovato, Trozzo,
  Trudeau, Tsai, Tsai, Tsang, Tsang, Tsao, Tse, Tso, Tsubono, Tsuchida,
  Tsukada, Tsuna, Tsutsui, Tsuzuki, Turconi, Tuyenbayev, Ubhi, Uchikata,
  Uchiyama, Udall, Ueda, Uehara, Ueno, Ueshima, Ugolini, Unnikrishnan,
  Uraguchi, Urban, Ushiba, Usman, Utina, Vahlbruch, Vajente, Vajpeyi, Valdes,
  Valentini, Valsan, van Bakel, van Beuzekom, van~den Brand, Van Den~Broeck,
  van Remortel, Vander-Hyde, van~der Schaaf, van Heijningen, van Putten,
  Vardaro, Vargas, Varma, Vas\'uth, Vecchio, Vedovato, Veitch, Veitch,
  Venkateswara, Venneberg, Venugopalan, Verkindt, Verma, Veske, Vetrano,
  Vicer\'e, Viets, Villa-Ortega, Vinet, Vitale, Vo, Vocca, von Reis, von
  Wrangel, Vorvick, Vyatchanin, Wade, Wade, Wagner, Walet, Walker, Wallace,
  Wallace, Walsh, Wang, Wang, Wang, Ward, Warner, Was, Washimi, Washington,
  Watchi, Weaver, Wei, Weinert, Weinstein, Weiss, Weller, Wellmann, Wen,
  We\ss{}els, Westhouse, Wette, Whelan, White, Whiting, Whittle, Wilken,
  Williams, Williams, Williamson, Willis, Willke, Wilson, Winkler, Wipf,
  Wlodarczyk, Woan, Woehler, Wofford, Wong, Wu, Wu, Wu, Wu, Wysocki, Xiao, Xu,
  Yamada, Yamamoto, Yamamoto, Yamamoto, Yamamoto, Yamashita, Yamazaki, Yang,
  Yang, Yang, Yang, Yang, Yap, Yeeles, Yelikar, Ying, Yokogawa, Yokoyama,
  Yokozawa, Yoon, Yoshioka, Yu, Yu, Yuzurihara, Zadro\ifmmode~\dot{z}\else
  \.{z}\fi{}ny, Zanolin, Zeidler, Zelenova, Zendri, Zevin, Zhan, Zhang, Zhang,
  Zhang, Zhang, Zhang, Zhao, Zhao, Zhao, Zhao, Zhou, Zhu, Zhu, Zucker, \&
  Zweizig}]{PhysRevD.104.022004}
Abbott, R., Abbott, T.~D., Abraham, S., {et~al.} 2021, Phys. Rev. D, 104,
  022004

\bibitem[{{Abbott} {et~al.}(2021){Abbott}, {Abbott}, {Acernese}, {Ackley},
  {Adams}, {Adhikari}, {Adhikari}, {Adya}, {Affeldt}, {Agarwal}, {Agathos},
  {Agatsuma}, {Aggarwal}, {Aguiar}, {Aiello}, {Ain}, {Ajith}, {Akcay},
  {Akutsu}, {Albanesi}, {Allocca}, {Altin}, {Amato}, {Anand}, {Anand},
  {Ananyeva}, {Anderson}, {Anderson}, {Ando}, {Andrade}, {Andres},
  {Andri{\'c}}, {Angelova}, {Ansoldi}, {Antelis}, {Antier}, {Appert}, {Arai},
  {Arai}, {Arai}, {Araki}, {Araya}, {Araya}, {Areeda}, {Ar{\`e}ne}, {Aritomi},
  {Arnaud}, {Arogeti}, {Aronson}, {Arun}, {Asada}, {Asali}, {Ashton}, {Aso},
  {Assiduo}, {Aston}, {Astone}, {Aubin}, {Austin}, {Babak}, {Badaracco},
  {Bader}, {Badger}, {Bae}, {Bae}, {Baer}, {Bagnasco}, {Bai}, {Baiotti},
  {Baird}, {Bajpai}, {Ball}, {Ballardin}, {Ballmer}, {Balsamo}, {Baltus},
  {Banagiri}, {Bankar}, {Barayoga}, {Barbieri}, {Barish}, {Barker}, {Barneo},
  {Barone}, {Barr}, {Barsotti}, {Barsuglia}, {Barta}, {Bartlett}, {Barton},
  {Bartos}, {Bassiri}, {Basti}, {Bawaj}, {Bayley}, {Baylor}, {Bazzan},
  {B{\'e}csy}, {Bedakihale}, {Bejger}, {Belahcene}, {Benedetto}, {Beniwal},
  {Bennett}, {Bentley}, {BenYaala}, {Bergamin}, {Berger}, {Bernuzzi}, {Berry},
  {Bersanetti}, {Bertolini}, {Betzwieser}, {Beveridge}, {Bhandare}, {Bhardwaj},
  {Bhattacharjee}, {Bhaumik}, {Bilenko}, {Billingsley}, {Bini}, {Birney},
  {Birnholtz}, {Biscans}, {Bischi}, {Biscoveanu}, {Bisht}, {Biswas}, {Bitossi},
  {Bizouard}, {Blackburn}, {Blair}, {Blair}, {Blair}, {Bobba}, {Bode}, {Boer},
  {Bogaert}, {Boldrini}, {Bonavena}, {Bondu}, {Bonilla}, {Bonnand}, {Booker},
  {Boom}, {Bork}, {Boschi}, {Bose}, {Bose}, {Bossilkov}, {Boudart},
  {Bouffanais}, {Bozzi}, {Bradaschia}, {Brady}, {Bramley}, {Branch},
  {Branchesi}, {Brandt}, {Brau}, {Breschi}, {Briant}, {Briggs}, {Brillet},
  {Brinkmann}, {Brockill}, {Brooks}, {Brooks}, {Brown}, {Brunett}, {Bruno},
  {Bruntz}, {Bryant}, {Bulik}, {Bulten}, {Buonanno}, {Buscicchio}, {Buskulic},
  {Buy}, {Byer}, {Cabourn Davies}, {Cadonati}, {Cagnoli}, {Cahillane},
  {Calder{\'o}n Bustillo}, {Callaghan}, {Callister}, {Calloni}, {Cameron},
  {Camp}, {Canepa}, {Canevarolo}, {Cannavacciuolo}, {Cannon}, {Cao}, {Cao},
  {Capocasa}, {Capote}, {Carapella}, {Carbognani}, {Carlin}, {Carney},
  {Carpinelli}, {Carrillo}, {Carullo}, {Carver}, {Casanueva Diaz}, {Casentini},
  {Castaldi}, {Caudill}, {Cavagli{\`a}}, {Cavalier}, {Cavalieri}, {Ceasar},
  {Cella}, {Cerd{\'a}-Dur{\'a}n}, {Cesarini}, {Chaibi}, {Chakravarti},
  {Chalathadka Subrahmanya}, {Champion}, {Chan}, {Chan}, {Chan}, {Chan},
  {Chan}, {Chandra}, {Chanial}, {Chao}, {Chapman-Bird}, {Charlton}, {Chase},
  {Chassande-Mottin}, {Chatterjee}, {Chatterjee}, {Chatterjee}, {Chaturvedi},
  {Chaty}, {Chatziioannou}, {Chen}, {Chen}, {Chen}, {Chen}, {Chen}, {Chen},
  {Chen}, {Chen}, {Cheng}, {Cheong}, {Cheung}, {Chia}, {Chiadini}, {Chiang},
  {Chiarini}, {Chierici}, {Chincarini}, {Chiofalo}, {Chiummo}, {Cho}, {Cho},
  {Choudhary}, {Choudhary}, {Christensen}, {Chu}, {Chu}, {Chu}, {Chua},
  {Chung}, {Ciani}, {Ciecielag}, {Cie{\'s}lar}, {Cifaldi}, {Ciobanu}, {Ciolfi},
  {Cipriano}, {Cirone}, {Clara}, {Clark}, {Clark}, {Clarke}, {Clearwater},
  {Clesse}, {Cleva}, {Coccia}, {Codazzo}, {Cohadon}, {Cohen}, {Cohen},
  {Colleoni}, {Collette}, {Colombo}, {Colpi}, {Compton}, {Constancio}, {Conti},
  {Cooper}, {Corban}, {Corbitt}, {Cordero-Carri{\'o}n}, {Corezzi}, {Corley},
  {Cornish}, {Corre}, {Corsi}, {Cortese}, {Costa}, {Cotesta}, {Coughlin},
  {Coulon}, {Countryman}, {Cousins}, {Couvares}, {Coward}, {Cowart}, {Coyne},
  {Coyne}, {Creighton}, {Creighton}, {Criswell}, {Croquette}, {Crowder},
  {Cudell}, {Cullen}, {Cumming}, {Cummings}, {Cunningham}, {Cuoco},
  {Cury{\l}o}, {Dabadie}, {Dal Canton}, {Dall'Osso}, {D{\'a}lya}, {Dana},
  {DaneshgaranBajastani}, {D'Angelo}, {Danila}, {Danilishin}, {D'Antonio},
  {Danzmann}, {Darsow-Fromm}, {Dasgupta}, {Datrier}, {Datta}, {Dattilo},
  {Dave}, {Davier}, {Davis}, {Davis}, {Daw}, {de Alarc{\'o}n}, {Dean}, {DeBra},
  {Deenadayalan}, {Degallaix}, {De Laurentis}, {Del{\'e}glise}, {Del Favero},
  {De Lillo}, {De Lillo}, {Del Pozzo}, {DeMarchi}, {De Matteis}, {D'Emilio},
  {Demos}, {Dent}, {Depasse}, {De Pietri}, {De Rosa}, {De Rossi}, {DeSalvo},
  {De Simone}, {Dhurandhar}, {D{\'\i}az}, {Diaz-Ortiz}, {Didio}, {Dietrich},
  {Di Fiore}, {Di Fronzo}, {Di Giorgio}, {Di Giovanni}, {Di Giovanni}, {Di
  Girolamo}, {Di Lieto}, {Ding}, {Di Pace}, {Di Palma}, {Di Renzo},
  {Divakarla}, {Dmitriev}, {Doctor}, {D'Onofrio}, {Donovan}, {Dooley},
  {Doravari}, {Dorrington}, {Drago}, {Driggers}, {Drori}, {Ducoin}, {Dupej},
  {Durante}, {D'Urso}, {Duverne}, {Dwyer}, {Eassa}, {Easter}, {Ebersold},
  {Eckhardt}, {Eddolls}, {Edelman}, {Edo}, {Edy}, {Effler}, {Eguchi},
  {Eichholz}, {Eikenberry}, {Eisenmann}, {Eisenstein}, {Ejlli}, {Engelby},
  {Enomoto}, {Errico}, {Essick}, {Estell{\'e}s}, {Estevez}, {Etienne}, {Etzel},
  {Evans}, {Evans}, {Ewing}, {Fafone}, {Fair}, {Fairhurst}, {Farah}, {Farinon},
  {Farr}, {Farr}, {Farrow}, {Fauchon-Jones}, {Favaro}, {Favata}, {Fays},
  {Fazio}, {Feicht}, {Fejer}, {Fenyvesi}, {Ferguson}, {Fernandez-Galiana},
  {Ferrante}, {Ferreira}, {Fidecaro}, {Figura}, {Fiori}, {Fishbach}, {Fisher},
  {Fittipaldi}, {Fiumara}, {Flaminio}, {Floden}, {Fong}, {Font}, {Fornal},
  {Forsyth}, {Franke}, {Frasca}, {Frasconi}, {Frederick}, {Freed}, {Frei},
  {Freise}, {Frey}, {Fritschel}, {Frolov}, {Fronz{\'e}}, {Fujii}, {Fujikawa},
  {Fukunaga}, {Fukushima}, {Fulda}, {Fyffe}, {Gabbard}, {Gabella}, {Gadre},
  {Gair}, {Gais}, {Galaudage}, {Gamba}, {Ganapathy}, {Ganguly}, {Gao},
  {Gaonkar}, {Garaventa}, {Garc{\'\i}a}, {Garc{\'\i}a-N{\'u}{\~n}ez},
  {Garc{\'\i}a-Quir{\'o}s}, {Garufi}, {Gateley}, {Gaudio}, {Gayathri}, {Ge},
  {Gemme}, {Gennai}, {George}, {George}, {Gerberding}, {Gergely}, {Gewecke},
  {Ghonge}, {Ghosh}, {Ghosh}, {Ghosh}, {Ghosh}, {Giacomazzo}, {Giacoppo},
  {Giaime}, {Giardina}, {Gibson}, {Gier}, {Giesler}, {Giri}, {Gissi},
  {Glanzer}, {Gleckl}, {Godwin}, {Goetz}, {Goetz}, {Gohlke}, {Golomb},
  {Goncharov}, {Gonz{\'a}lez}, {Gopakumar}, {Gosselin}, {Gouaty}, {Gould},
  {Grace}, {Grado}, {Granata}, {Granata}, {Grant}, {Gras}, {Grassia}, {Gray},
  {Gray}, {Greco}, {Green}, {Green}, {Gretarsson}, {Gretarsson}, {Griffith},
  {Griffiths}, {Griggs}, {Grignani}, {Grimaldi}, {Grimm}, {Grote}, {Grunewald},
  {Gruning}, {Guerra}, {Guidi}, {Guimaraes}, {Guix{\'e}}, {Gulati}, {Guo},
  {Guo}, {Gupta}, {Gupta}, {Gupta}, {Gustafson}, {Gustafson}, {Guzman}, {Ha},
  {Haegel}, {Hagiwara}, {Haino}, {Halim}, {Hall}, {Hamilton}, {Hammond}, {Han},
  {Haney}, {Hanks}, {Hanna}, {Hannam}, {Hannuksela}, {Hansen}, {Hansen},
  {Hanson}, {Harder}, {Hardwick}, {Haris}, {Harms}, {Harry}, {Harry},
  {Hartwig}, {Hasegawa}, {Haskell}, {Hasskew}, {Haster}, {Hattori}, {Haughian},
  {Hayakawa}, {Hayama}, {Hayes}, {Healy}, {Heidmann}, {Heidt}, {Heintze},
  {Heinze}, {Heinzel}, {Heitmann}, {Hellman}, {Hello}, {Helmling-Cornell},
  {Hemming}, {Hendry}, {Heng}, {Hennes}, {Hennig}, {Hennig}, {Hernandez},
  {Hernandez Vivanco}, {Heurs}, {Hild}, {Hill}, {Himemoto}, {Hines},
  {Hiranuma}, {Hirata}, {Hirose}, {Hochheim}, {Hofman}, {Hohmann}, {Holcomb},
  {Holland}, {Holley-Bockelmann}, {Hollows}, {Holmes}, {Holt}, {Holz}, {Hong},
  {Hopkins}, {Hough}, {Hourihane}, {Howell}, {Hoy}, {Hoyland}, {Hreibi},
  {Hsieh}, {Hsu}, {Huang}, {Huang}, {Huang}, {Huang}, {Huang}, {Huang},
  {H{\"u}bner}, {Huddart}, {Hughey}, {Hui}, {Hui}, {Husa}, {Huttner},
  {Huxford}, {Huynh-Dinh}, {Ide}, {Idzkowski}, {Iess}, {Ikenoue}, {Imam},
  {Inayoshi}, {Ingram}, {Inoue}, {Ioka}, {Isi}, {Isleif}, {Ito}, {Itoh},
  {Iyer}, {Izumi}, {JaberianHamedan}, {Jacqmin}, {Jadhav}, {Jadhav}, {James},
  {Jan}, {Jani}, {Janquart}, {Janssens}, {Janthalur}, {Jaranowski}, {Jariwala},
  {Jaume}, {Jenkins}, {Jenner}, {Jeon}, {Jeunon}, {Jia}, {Jin}, {Johns},
  {Johnson-McDaniel}, {Jones}, {Jones}, {Jones}, {Jones}, {Jones}, {Jonker},
  {Ju}, {Jung}, {Jung}, {Junker}, {Juste}, {Kaihotsu}, {Kajita}, {Kakizaki},
  {Kalaghatgi}, {Kalogera}, {Kamai}, {Kamiizumi}, {Kanda}, {Kandhasamy},
  {Kang}, {Kanner}, {Kao}, {Kapadia}, {Kapasi}, {Karat}, {Karathanasis},
  {Karki}, {Kashyap}, {Kasprzack}, {Kastaun}, {Katsanevas}, {Katsavounidis},
  {Katzman}, {Kaur}, {Kawabe}, {Kawaguchi}, {Kawai}, {Kawasaki},
  {K{\'e}f{\'e}lian}, {Keitel}, {Key}, {Khadka}, {Khalili}, {Khan}, {Khazanov},
  {Khetan}, {Khursheed}, {Kijbunchoo}, {Kim}, {Kim}, {Kim}, {Kim}, {Kim},
  {Kim}, {Kimball}, {Kimura}, {Kinley-Hanlon}, {Kirchhoff}, {Kissel}, {Kita},
  {Kitazawa}, {Kleybolte}, {Klimenko}, {Knee}, {Knowles}, {Knyazev}, {Koch},
  {Koekoek}, {Kojima}, {Kokeyama}, {Koley}, {Kolitsidou}, {Kolstein}, {Komori},
  {Kondrashov}, {Kong}, {Kontos}, {Koper}, {Korobko}, {Kotake}, {Kovalam},
  {Kozak}, {Kozakai}, {Kozu}, {Kringel}, {Krishnendu}, {Kr{\'o}lak}, {Kuehn},
  {Kuei}, {Kuijer}, {Kulkarni}, {Kumar}, {Kumar}, {Kumar}, {Kumar}, {Kume},
  {Kuns}, {Kuo}, {Kuo}, {Kuromiya}, {Kuroyanagi}, {Kusayanagi}, {Kuwahara},
  {Kwak}, {Lagabbe}, {Laghi}, {Lalande}, {Lam}, {Lamberts}, {Landry}, {Lane},
  {Lang}, {Lange}, {Lantz}, {La Rosa}, {Lartaux-Vollard}, {Lasky}, {Laxen},
  {Lazzarini}, {Lazzaro}, {Leaci}, {Leavey}, {Lecoeuche}, {Lee}, {Lee}, {Lee},
  {Lee}, {Lee}, {Lee}, {Lehmann}, {Lema{\^\i}tre}, {Leonardi}, {Leroy},
  {Letendre}, {Levesque}, {Levin}, {Leviton}, {Leyde}, {Li}, {Li}, {Li}, {Li},
  {Li}, {Li}, {Lin}, {Lin}, {Lin}, {Lin}, {Lin}, {Linde}, {Linker}, {Linley},
  {Littenberg}, {Liu}, {Liu}, {Liu}, {Liu}, {Llamas}, {Llorens-Monteagudo},
  {Lo}, {Lockwood}, {Loh}, {London}, {Longo}, {Lopez}, {Lopez Portilla},
  {Lorenzini}, {Loriette}, {Lormand}, {Losurdo}, {Lott}, {Lough}, {Lousto},
  {Lovelace}, {Lucaccioni}, {L{\"u}ck}, {Lumaca}, {Lundgren}, {Luo}, {Lynam},
  {Macas}, {MacInnis}, {Macleod}, {MacMillan}, {Macquet}, {Maga{\~n}a
  Hernandez}, {Magazz{\`u}}, {Magee}, {Maggiore}, {Magnozzi}, {Mahesh},
  {Majorana}, {Makarem}, {Maksimovic}, {Maliakal}, {Malik}, {Man}, {Mandic},
  {Mangano}, {Mango}, {Mansell}, {Manske}, {Mantovani}, {Mapelli},
  {Marchesoni}, {Marchio}, {Marion}, {Mark}, {M{\'a}rka}, {M{\'a}rka},
  {Markakis}, {Markosyan}, {Markowitz}, {Maros}, {Marquina}, {Marsat},
  {Martelli}, {Martin}, {Martin}, {Martinez}, {Martinez}, {Martinez},
  {Martinovic}, {Martynov}, {Marx}, {Masalehdan}, {Mason}, {Massera},
  {Masserot}, {Massinger}, {Masso-Reid}, {Mastrogiovanni}, {Matas},
  {Mateu-Lucena}, {Matichard}, {Matiushechkina}, {Mavalvala}, {McCann},
  {McCarthy}, {McClelland}, {McClincy}, {McCormick}, {McCuller}, {McGhee},
  {McGuire}, {McIsaac}, {McIver}, {McRae}, {McWilliams}, {Meacher}, {Mehmet},
  {Mehta}, {Meijer}, {Melatos}, {Melchor}, {Mendell}, {Menendez-Vazquez},
  {Menoni}, {Mercer}, {Mereni}, {Merfeld}, {Merilh}, {Merritt}, {Merzougui},
  {Meshkov}, {Messenger}, {Messick}, {Meyers}, {Meylahn}, {Mhaske}, {Miani},
  {Miao}, {Michaloliakos}, {Michel}, {Michimura}, {Middleton}, {Milano},
  {Miller}, {Miller}, {Miller}, {Millhouse}, {Mills}, {Milotti}, {Minazzoli},
  {Minenkov}, {Mio}, {Mir}, {Miravet-Ten{\'e}s}, {Mishra}, {Mishra}, {Mistry},
  {Mitra}, {Mitrofanov}, {Mitselmakher}, {Mittleman}, {Miyakawa}, {Miyamoto},
  {Miyazaki}, {Miyo}, {Miyoki}, {Mo}, {Modafferi}, {Moguel}, {Mogushi},
  {Mohapatra}, {Mohite}, {Molina}, {Molina-Ruiz}, {Mondin}, {Montani}, {Moore},
  {Moraru}, {Morawski}, {More}, {Moreno}, {Moreno}, {Mori}, {Morisaki},
  {Moriwaki}, {Morr{\'a}s}, {Mours}, {Mow-Lowry}, {Mozzon}, {Muciaccia},
  {Mukherjee}, {Mukherjee}, {Mukherjee}, {Mukherjee}, {Mukherjee}, {Mukund},
  {Mullavey}, {Munch}, {Mu{\~n}iz}, {Murray}, {Musenich}, {Muusse}, {Nadji},
  {Nagano}, {Nagano}, {Nagar}, {Nakamura}, {Nakano}, {Nakano}, {Nakashima},
  {Nakayama}, {Napolano}, {Nardecchia}, {Narikawa}, {Naticchioni}, {Nayak},
  {Nayak}, {Negishi}, {Neil}, {Neilson}, {Nelemans}, {Nelson}, {Nery},
  {Neubauer}, {Neunzert}, {Ng}, {Ng}, {Nguyen}, {Nguyen}, {Nguyen}, {Nguyen
  Quynh}, {Ni}, {Nichols}, {Nishizawa}, {Nissanke}, {Nitoglia}, {Nocera},
  {Norman}, {North}, {Nozaki}, {Nu{\~n}o Siles}, {Nuttall}, {Oberling},
  {O'Brien}, {Obuchi}, {O'Dell}, {Oelker}, {Ogaki}, {Oganesyan}, {Oh}, {Oh},
  {Oh}, {Ohashi}, {Ohishi}, {Ohkawa}, {Ohme}, {Ohta}, {Okada}, {Okutani},
  {Okutomi}, {Olivetto}, {Oohara}, {Ooi}, {Oram}, {O'Reilly}, {Ormiston},
  {Ormsby}, {Ortega}, {O'Shaughnessy}, {O'Shea}, {Oshino}, {Ossokine},
  {Osthelder}, {Otabe}, {Ottaway}, {Overmier}, {Pace}, {Pagano}, {Page},
  {Pagliaroli}, {Pai}, {Pai}, {Palamos}, {Palashov}, {Palomba}, {Pan}, {Pan},
  {Panda}, {Pang}, {Pang}, {Pankow}, {Pannarale}, {Pant}, {Panther},
  {Paoletti}, {Paoli}, {Paolone}, {Parisi}, {Park}, {Park}, {Parker},
  {Pascucci}, {Pasqualetti}, {Passaquieti}, {Passuello}, {Patel}, {Pathak},
  {Patricelli}, {Patron}, {Paul}, {Payne}, {Pedraza}, {Pegoraro}, {Pele},
  {Pe{\~n}a Arellano}, {Penn}, {Perego}, {Pereira}, {Pereira}, {Perez},
  {P{\'e}rigois}, {Perkins}, {Perreca}, {Perri{\`e}s}, {Petermann},
  {Petterson}, {Pfeiffer}, {Pham}, {Phukon}, {Piccinni}, {Pichot},
  {Piendibene}, {Piergiovanni}, {Pierini}, {Pierro}, {Pillant}, {Pillas},
  {Pilo}, {Pinard}, {Pinto}, {Pinto}, {Piotrzkowski}, {Piotrzkowski},
  {Pirello}, {Pitkin}, {Placidi}, {Planas}, {Plastino}, {Pluchar}, {Poggiani},
  {Polini}, {Pong}, {Ponrathnam}, {Popolizio}, {Porter}, {Poulton}, {Powell},
  {Pracchia}, {Pradier}, {Prajapati}, {Prasai}, {Prasanna}, {Pratten},
  {Principe}, {Prodi}, {Prokhorov}, {Prosposito}, {Prudenzi}, {Puecher},
  {Punturo}, {Puosi}, {Puppo}, {P{\"u}rrer}, {Qi}, {Quetschke},
  {Quitzow-James}, {Qutob}, {Raab}, {Raaijmakers}, {Radkins}, {Radulesco},
  {Raffai}, {Rail}, {Raja}, {Rajan}, {Ramirez}, {Ramirez}, {Ramos-Buades},
  {Rana}, {Rapagnani}, {Rapol}, {Ray}, {Raymond}, {Raza}, {Razzano}, {Read},
  {Rees}, {Regimbau}, {Rei}, {Reid}, {Reid}, {Reitze}, {Relton}, {Renzini},
  {Rettegno}, {Reza}, {Rezac}, {Ricci}, {Richards}, {Richardson}, {Richardson},
  {Riemenschneider}, {Riles}, {Rinaldi}, {Rink}, {Rizzo}, {Robertson}, {Robie},
  {Robinet}, {Rocchi}, {Rodriguez}, {Rolland}, {Rollins}, {Romanelli},
  {Romano}, {Romel}, {Romero-Rodr{\'\i}guez}, {Romero-Shaw}, {Romie},
  {Ronchini}, {Rosa}, {Rose}, {Rosi{\'n}ska}, {Ross}, {Rowan}, {Rowlinson},
  {Roy}, {Roy}, {Roy}, {Rozza}, {Ruggi}, {Ruiz-Rocha}, {Ryan}, {Sachdev},
  {Sadecki}, {Sadiq}, {Sago}, {Saito}, {Saito}, {Sakai}, {Sakai},
  {Sakellariadou}, {Sakuno}, {Salafia}, {Salconi}, {Saleem}, {Salemi},
  {Samajdar}, {Sanchez}, {Sanchez}, {Sanchez}, {Sanchis-Gual}, {Sanders},
  {Sanuy}, {Saravanan}, {Sarin}, {Sassolas}, {Satari}, {Sathyaprakash}, {Sato},
  {Sato}, {Sauter}, {Savage}, {Sawada}, {Sawant}, {Sawant}, {Sayah},
  {Schaetzl}, {Scheel}, {Scheuer}, {Schiworski}, {Schmidt}, {Schmidt},
  {Schnabel}, {Schneewind}, {Schofield}, {Sch{\"o}nbeck}, {Schulte}, {Schutz},
  {Schwartz}, {Scott}, {Scott}, {Seglar-Arroyo}, {Sekiguchi}, {Sekiguchi},
  {Sellers}, {Sengupta}, {Sentenac}, {Seo}, {Sequino}, {Sergeev}, {Setyawati},
  {Shaffer}, {Shahriar}, {Shams}, {Shao}, {Sharma}, {Sharma}, {Shawhan},
  {Shcheblanov}, {Shibagaki}, {Shikauchi}, {Shimizu}, {Shimoda}, {Shimode},
  {Shinkai}, {Shishido}, {Shoda}, {Shoemaker}, {Shoemaker}, {ShyamSundar},
  {Sieniawska}, {Sigg}, {Singer}, {Singh}, {Singh}, {Singha}, {Sintes},
  {Sipala}, {Skliris}, {Slagmolen}, {Slaven-Blair}, {Smetana}, {Smith},
  {Smith}, {Soldateschi}, {Somala}, {Somiya}, {Son}, {Soni}, {Soni}, {Sordini},
  {Sorrentino}, {Sorrentino}, {Sotani}, {Soulard}, {Souradeep}, {Sowell},
  {Spagnuolo}, {Spencer}, {Spera}, {Srinivasan}, {Srivastava}, {Srivastava},
  {Staats}, {Stachie}, {Steer}, {Steinhoff}, {Steinlechner}, {Steinlechner},
  {Stevenson}, {Stops}, {Stover}, {Strain}, {Strang}, {Stratta}, {Strunk},
  {Sturani}, {Stuver}, {Sudhagar}, {Sudhir}, {Sugimoto}, {Suh}, {Sullivan},
  {Sullivan}, {Summerscales}, {Sun}, {Sun}, {Sunil}, {Sur}, {Suresh}, {Sutton},
  {Suzuki}, {Suzuki}, {Swinkels}, {Szczepa{\'n}czyk}, {Szewczyk}, {Tacca},
  {Tagoshi}, {Tait}, {Takahashi}, {Takahashi}, {Takamori}, {Takano}, {Takeda},
  {Takeda}, {Talbot}, {Talbot}, {Tanaka}, {Tanaka}, {Tanaka}, {Tanaka},
  {Tanaka}, {Tanasijczuk}, {Tanioka}, {Tanner}, {Tao}, {Tao}, {Tapia San
  Mart{\'\i}n}, {Taranto}, {Tasson}, {Telada}, {Tenorio}, {Terhune},
  {Terkowski}, {Thirugnanasambandam}, {Thomas}, {Thomas}, {Thomas}, {Thompson},
  {Thondapu}, {Thorne}, {Thrane}, {Tiwari}, {Tiwari}, {Tiwari}, {Toivonen},
  {Toland}, {Tolley}, {Tomaru}, {Tomigami}, {Tomura}, {Tonelli},
  {Torres-Forn{\'e}}, {Torrie}, {Tosta e Melo}, {T{\"o}yr{\"a}}, {Trapananti},
  {Travasso}, {Traylor}, {Trevor}, {Tringali}, {Tripathee}, {Troiano},
  {Trovato}, {Trozzo}, {Trudeau}, {Tsai}, {Tsai}, {Tsang}, {Tsang}, {Tsao},
  {Tse}, {Tso}, {Tsubono}, {Tsuchida}, {Tsukada}, {Tsuna}, {Tsutsui},
  {Tsuzuki}, {Turbang}, {Turconi}, {Tuyenbayev}, {Ubhi}, {Uchikata},
  {Uchiyama}, {Udall}, {Ueda}, {Uehara}, {Ueno}, {Ueshima}, {Unnikrishnan},
  {Uraguchi}, {Urban}, {Ushiba}, {Utina}, {Vahlbruch}, {Vajente}, {Vajpeyi},
  {Valdes}, {Valentini}, {Valsan}, {van Bakel}, {van Beuzekom}, {van den
  Brand}, {Van Den Broeck}, {Vander-Hyde}, {van der Schaaf}, {van Heijningen},
  {Vanosky}, {van Putten}, {van Remortel}, {Vardaro}, {Vargas}, {Varma},
  {Vas{\'u}th}, {Vecchio}, {Vedovato}, {Veitch}, {Veitch}, {Venneberg},
  {Venugopalan}, {Verkindt}, {Verma}, {Verma}, {Veske}, {Vetrano},
  {Vicer{\'e}}, {Vidyant}, {Viets}, {Vijaykumar}, {Villa-Ortega}, {Vinet},
  {Virtuoso}, {Vitale}, {Vo}, {Vocca}, {von Reis}, {von Wrangel}, {Vorvick},
  {Vyatchanin}, {Wade}, {Wade}, {Wagner}, {Walet}, {Walker}, {Wallace},
  {Wallace}, {Walsh}, {Wang}, {Wang}, {Wang}, {Ward}, {Warner}, {Was},
  {Washimi}, {Washington}, {Watchi}, {Weaver}, {Webster}, {Weinert},
  {Weinstein}, {Weiss}, {Weller}, {Weller}, {Wellmann}, {Wen}, {We{\ss}els},
  {Wette}, {Whelan}, {White}, {Whiting}, {Whittle}, {Wilken}, {Williams},
  {Williams}, {Williams}, {Williamson}, {Willis}, {Willke}, {Wilson},
  {Winkler}, {Wipf}, {Wlodarczyk}, {Woan}, {Woehler}, {Wofford}, {Wong}, {Wu},
  {Wu}, {Wu}, {Wu}, {Wysocki}, {Xiao}, {Xu}, {Yamada}, {Yamamoto}, {Yamamoto},
  {Yamamoto}, {Yamamoto}, {Yamashita}, {Yamazaki}, {Yang}, {Yang}, {Yang},
  {Yang}, {Yang}, {Yap}, {Yeeles}, {Yelikar}, {Ying}, {Yokogawa}, {Yokoyama},
  {Yokozawa}, {Yoo}, {Yoshioka}, {Yu}, {Yu}, {Yuzurihara}, {Zadro{\.z}ny},
  {Zanolin}, {Zeidler}, {Zelenova}, {Zendri}, {Zevin}, {Zhan}, {Zhang},
  {Zhang}, {Zhang}, {Zhang}, {Zhang}, {Zhao}, {Zhao}, {Zhao}, {Zhao}, {Zheng},
  {Zhou}, {Zhou}, {Zhu}, {Zhu}, {Zimmerman}, {Zlochower}, {Zucker}, \&
  {Zweizig}}]{2021arXiv211103606T}
{Abbott}, R., {Abbott}, T.~D., {Acernese}, F., {et~al.} 2021, arXiv e-prints,
  arXiv:2111.03606

\bibitem[{{Abbott} {et~al.}(2023){Abbott}, {Abbott}, {Acernese}, {Ackley},
  {Adams}, {Adhikari}, {Adhikari}, {Adya}, {Affeldt}, {Agarwal}, {Agathos},
  {Agatsuma}, {Aggarwal}, {Aguiar}, {Aiello}, {Ain}, {Ajith}, {Akutsu}, {de
  Alarc{\'o}n}, {Akcay}, {Albanesi}, {Allocca}, {Altin}, {Amato}, {Anand},
  {Anand}, {Ananyeva}, {Anderson}, {Anderson}, {Ando}, {Andrade}, {Andres},
  {Andri{\'c}}, {Angelova}, {Ansoldi}, {Antelis}, {Antier}, {Antonini},
  {Appert}, {Arai}, {Arai}, {Arai}, {Araki}, {Araya}, {Araya}, {Areeda},
  {Ar{\`e}ne}, {Aritomi}, {Arnaud}, {Arogeti}, {Aronson}, {Arun}, {Asada},
  {Asali}, {Ashton}, {Aso}, {Assiduo}, {Aston}, {Astone}, {Aubin}, {Austin},
  {Babak}, {Badaracco}, {Bader}, {Badger}, {Bae}, {Bae}, {Baer}, {Bagnasco},
  {Bai}, {Baiotti}, {Baird}, {Bajpai}, {Ball}, {Ballardin}, {Ballmer},
  {Balsamo}, {Baltus}, {Banagiri}, {Bankar}, {Barayoga}, {Barbieri}, {Barish},
  {Barker}, {Barneo}, {Barone}, {Barr}, {Barsotti}, {Barsuglia}, {Barta},
  {Bartlett}, {Barton}, {Bartos}, {Bassiri}, {Basti}, {Bawaj}, {Bayley},
  {Baylor}, {Bazzan}, {B{\'e}csy}, {Bedakihale}, {Bejger}, {Belahcene},
  {Benedetto}, {Beniwal}, {Bennett}, {Bentley}, {Benyaala}, {Bergamin},
  {Berger}, {Bernuzzi}, {Berry}, {Bersanetti}, {Bertolini}, {Betzwieser},
  {Beveridge}, {Bhandare}, {Bhardwaj}, {Bhattacharjee}, {Bhaumik}, {Bilenko},
  {Billingsley}, {Bini}, {Birney}, {Birnholtz}, {Biscans}, {Bischi},
  {Biscoveanu}, {Bisht}, {Biswas}, {Bitossi}, {Bizouard}, {Blackburn}, {Blair},
  {Blair}, {Blair}, {Bobba}, {Bode}, {Boer}, {Bogaert}, {Boldrini}, {Bonavena},
  {Bondu}, {Bonilla}, {Bonnand}, {Booker}, {Boom}, {Bork}, {Boschi}, {Bose},
  {Bose}, {Bossilkov}, {Boudart}, {Bouffanais}, {Bozzi}, {Bradaschia}, {Brady},
  {Bramley}, {Branch}, {Branchesi}, {Brandt}, {Brau}, {Breschi}, {Briant},
  {Briggs}, {Brillet}, {Brinkmann}, {Brockill}, {Brooks}, {Brooks}, {Brown},
  {Brunett}, {Bruno}, {Bruntz}, {Bryant}, {Bulik}, {Bulten}, {Buonanno},
  {Buscicchio}, {Buskulic}, {Buy}, {Byer}, {Cadonati}, {Cagnoli}, {Cahillane},
  {Bustillo}, {Callaghan}, {Callister}, {Calloni}, {Cameron}, {Camp}, {Canepa},
  {Canevarolo}, {Cannavacciuolo}, {Cannon}, {Cao}, {Cao}, {Capocasa}, {Capote},
  {Carapella}, {Carbognani}, {Carlin}, {Carney}, {Carpinelli}, {Carrillo},
  {Carullo}, {Carver}, {Diaz}, {Casentini}, {Castaldi}, {Caudill},
  {Cavagli{\`a}}, {Cavalier}, {Cavalieri}, {Ceasar}, {Cella},
  {Cerd{\'a}-Dur{\'a}n}, {Cesarini}, {Chaibi}, {Chakravarti}, {Subrahmanya},
  {Champion}, {Chan}, {Chan}, {Chan}, {Chan}, {Chan}, {Chandra}, {Chanial},
  {Chao}, {Chapman-Bird}, {Charlton}, {Chase}, {Chassande-Mottin},
  {Chatterjee}, {Chatterjee}, {Chatterjee}, {Chaturvedi}, {Chaty},
  {Chatziioannou}, {Chen}, {Chen}, {Chen}, {Chen}, {Chen}, {Chen}, {Chen},
  {Chen}, {Cheng}, {Cheong}, {Cheung}, {Chia}, {Chiadini}, {Chiang},
  {Chiarini}, {Chierici}, {Chincarini}, {Chiofalo}, {Chiummo}, {Cho}, {Cho},
  {Choudhary}, {Choudhary}, {Christensen}, {Chu}, {Chu}, {Chu}, {Chua},
  {Chung}, {Ciani}, {Ciecielag}, {Cie{\'s}lar}, {Cifaldi}, {Ciobanu}, {Ciolfi},
  {Cipriano}, {Cirone}, {Clara}, {Clark}, {Clark}, {Clarke}, {Clearwater},
  {Clesse}, {Cleva}, {Coccia}, {Codazzo}, {Cohadon}, {Cohen}, {Cohen},
  {Colleoni}, {Collette}, {Colombo}, {Colpi}, {Compton}, {Constancio}, {Conti},
  {Cooper}, {Corban}, {Corbitt}, {Cordero-Carri{\'o}n}, {Corezzi}, {Corley},
  {Cornish}, {Corre}, {Corsi}, {Cortese}, {Costa}, {Cotesta}, {Coughlin},
  {Coulon}, {Countryman}, {Cousins}, {Couvares}, {Coward}, {Cowart}, {Coyne},
  {Coyne}, {Creighton}, {Creighton}, {Criswell}, {Croquette}, {Crowder},
  {Cudell}, {Cullen}, {Cumming}, {Cummings}, {Cunningham}, {Cuoco},
  {Cury{\l}o}, {Dabadie}, {Canton}, {Dall'Osso}, {D{\'a}lya}, {Dana},
  {Daneshgaranbajastani}, {D'Angelo}, {Danila}, {Danilishin}, {D'Antonio},
  {Danzmann}, {Darsow-Fromm}, {Dasgupta}, {Datrier}, {Datta}, {Dattilo},
  {Dave}, {Davier}, {Davies}, {Davis}, {Davis}, {Daw}, {Dean}, {Debra},
  {Deenadayalan}, {Degallaix}, {de Laurentis}, {Del{\'e}glise}, {Del Favero},
  {de Lillo}, {de Lillo}, {Del Pozzo}, {Demarchi}, {de Matteis}, {D'Emilio},
  {Demos}, {Dent}, {Depasse}, {de Pietri}, {De Rosa}, {de Rossi}, {Desalvo},
  {de Simone}, {Dhurandhar}, {D{\'\i}az}, {Diaz-Ortiz}, {Didio}, {Dietrich},
  {di Fiore}, {di Fronzo}, {di Giorgio}, {di Giovanni}, {di Giovanni}, {di
  Girolamo}, {di Lieto}, {Ding}, {di Pace}, {di Palma}, {di Renzo},
  {Divakarla}, {Dmitriev}, {Doctor}, {D'Onofrio}, {Donovan}, {Dooley},
  {Doravari}, {Dorrington}, {Drago}, {Driggers}, {Drori}, {Ducoin}, {Dupej},
  {Durante}, {D'Urso}, {Duverne}, {Dwyer}, {Eassa}, {Easter}, {Ebersold},
  {Eckhardt}, {Eddolls}, {Edelman}, {Edo}, {Edy}, {Effler}, {Eguchi},
  {Eichholz}, {Eikenberry}, {Eisenmann}, {Eisenstein}, {Ejlli}, {Engelby},
  {Enomoto}, {Errico}, {Essick}, {Estell{\'e}s}, {Estevez}, {Etienne}, {Etzel},
  {Evans}, {Evans}, {Ewing}, {Fafone}, {Fair}, {Fairhurst}, {Farah}, {Farinon},
  {Farr}, {Farr}, {Farrow}, {Fauchon-Jones}, {Favaro}, {Favata}, {Fays},
  {Fazio}, {Feicht}, {Fejer}, {Fenyvesi}, {Ferguson}, {Fernandez-Galiana},
  {Ferrante}, {Ferreira}, {Fidecaro}, {Figura}, {Fiori}, {Fishbach}, {Fisher},
  {Fittipaldi}, {Fiumara}, {Flaminio}, {Floden}, {Fong}, {Font}, {Fornal},
  {Forsyth}, {Franke}, {Frasca}, {Frasconi}, {Frederick}, {Freed}, {Frei},
  {Freise}, {Frey}, {Fritschel}, {Frolov}, {Fronz{\'e}}, {Fujii}, {Fujikawa},
  {Fukunaga}, {Fukushima}, {Fulda}, {Fyffe}, {Gabbard}, {Gadre}, {Gair},
  {Gais}, {Galaudage}, {Gamba}, {Ganapathy}, {Ganguly}, {Gao}, {Gaonkar},
  {Garaventa}, {Garc{\'\i}a}, {Garc{\'\i}a-N{\'u}{\~n}ez},
  {Garc{\'\i}a-Quir{\'o}s}, {Garufi}, {Gateley}, {Gaudio}, {Gayathri}, {Ge},
  {Gemme}, {Gennai}, {George}, {George}, {Gerberding}, {Gergely}, {Gewecke},
  {Ghonge}, {Ghosh}, {Ghosh}, {Ghosh}, {Ghosh}, {Giacomazzo}, {Giacoppo},
  {Giaime}, {Giardina}, {Gibson}, {Gier}, {Giesler}, {Giri}, {Gissi},
  {Glanzer}, {Gleckl}, {Godwin}, {Golomb}, {Goetz}, {Goetz}, {Gohlke},
  {Goncharov}, {Gonz{\'a}lez}, {Gopakumar}, {Gosselin}, {Gouaty}, {Gould},
  {Grace}, {Grado}, {Granata}, {Granata}, {Grant}, {Gras}, {Grassia}, {Gray},
  {Gray}, {Greco}, {Green}, {Green}, {Gretarsson}, {Gretarsson}, {Griffith},
  {Griffiths}, {Griggs}, {Grignani}, {Grimaldi}, {Grimm}, {Grote}, {Grunewald},
  {Gruning}, {Guerra}, {Guidi}, {Guimaraes}, {Guix{\'e}}, {Gulati}, {Guo},
  {Guo}, {Gupta}, {Gupta}, {Gupta}, {Gustafson}, {Gustafson}, {Guzman}, {Ha},
  {Haegel}, {Hagiwara}, {Haino}, {Halim}, {Hall}, {Hamilton}, {Hammond}, {Han},
  {Haney}, {Hanks}, {Hanna}, {Hannam}, {Hannuksela}, {Hansen}, {Hansen},
  {Hanson}, {Harder}, {Hardwick}, {Haris}, {Harms}, {Harry}, {Harry},
  {Hartwig}, {Hasegawa}, {Haskell}, {Hasskew}, {Haster}, {Hattori}, {Haughian},
  {Hayakawa}, {Hayama}, {Hayes}, {Healy}, {Heidmann}, {Heidt}, {Heintze},
  {Heinze}, {Heinzel}, {Heitmann}, {Hellman}, {Hello}, {Helmling-Cornell},
  {Hemming}, {Hendry}, {Heng}, {Hennes}, {Hennig}, {Hennig}, {Hernandez},
  {Vivanco}, {Heurs}, {Hild}, {Hill}, {Himemoto}, {Hines}, {Hiranuma},
  {Hirata}, {Hirose}, {Hochheim}, {Hofman}, {Hohmann}, {Holcomb}, {Holland},
  {Hollows}, {Holmes}, {Holt}, {Holz}, {Hong}, {Hopkins}, {Hough}, {Hourihane},
  {Howell}, {Hoy}, {Hoyland}, {Hreibi}, {Hsieh}, {Hsu}, {Huang}, {Huang},
  {Huang}, {Huang}, {Huang}, {Huang}, {H{\"u}bner}, {Huddart}, {Hughey}, {Hui},
  {Hui}, {Husa}, {Huttner}, {Huxford}, {Huynh-Dinh}, {Ide}, {Idzkowski},
  {Iess}, {Ikenoue}, {Imam}, {Inayoshi}, {Ingram}, {Inoue}, {Ioka}, {Isi},
  {Isleif}, {Ito}, {Itoh}, {Iyer}, {Izumi}, {Jaberianhamedan}, {Jacqmin},
  {Jadhav}, {Jadhav}, {James}, {Jan}, {Jani}, {Janquart}, {Janssens},
  {Janthalur}, {Jaranowski}, {Jariwala}, {Jaume}, {Jenkins}, {Jenner}, {Jeon},
  {Jeunon}, {Jia}, {Jin}, {Johns}, {Jones}, {Jones}, {Jones}, {Jones}, {Jones},
  {Jonker}, {Ju}, {Jung}, {Jung}, {Junker}, {Juste}, {Kaihotsu}, {Kajita},
  {Kakizaki}, {Kalaghatgi}, {Kalogera}, {Kamai}, {Kamiizumi}, {Kanda},
  {Kandhasamy}, {Kang}, {Kanner}, {Kao}, {Kapadia}, {Kapasi}, {Karat},
  {Karathanasis}, {Karki}, {Kashyap}, {Kasprzack}, {Kastaun}, {Katsanevas},
  {Katsavounidis}, {Katzman}, {Kaur}, {Kawabe}, {Kawaguchi}, {Kawai},
  {Kawasaki}, {K{\'e}f{\'e}lian}, {Keitel}, {Key}, {Khadka}, {Khalili}, {Khan},
  {Khazanov}, {Khetan}, {Khursheed}, {Kijbunchoo}, {Kim}, {Kim}, {Kim}, {Kim},
  {Kim}, {Kim}, {Kimball}, {Kimura}, {Kinley-Hanlon}, {Kirchhoff}, {Kissel},
  {Kita}, {Kitazawa}, {Kleybolte}, {Klimenko}, {Knee}, {Knowles}, {Knyazev},
  {Koch}, {Koekoek}, {Kojima}, {Kokeyama}, {Koley}, {Kolitsidou}, {Kolstein},
  {Komori}, {Kondrashov}, {Kong}, {Kontos}, {Koper}, {Korobko}, {Kotake},
  {Kovalam}, {Kozak}, {Kozakai}, {Kozu}, {Kringel}, {Krishnendu}, {Kr{\'o}lak},
  {Kuehn}, {Kuei}, {Kuijer}, {Kulkarni}, {Kumar}, {Kumar}, {Kumar}, {Kumar},
  {Kume}, {Kuns}, {Kuo}, {Kuo}, {Kuromiya}, {Kuroyanagi}, {Kusayanagi},
  {Kuwahara}, {Kwak}, {Lagabbe}, {Laghi}, {Lalande}, {Lam}, {Lamberts},
  {Landry}, {Landry}, {Lane}, {Lang}, {Lange}, {Lantz}, {La Rosa},
  {Lartaux-Vollard}, {Lasky}, {Laxen}, {Lazzarini}, {Lazzaro}, {Leaci},
  {Leavey}, {Lecoeuche}, {Lee}, {Lee}, {Lee}, {Lee}, {Lee}, {Lee}, {Lehmann},
  {Lema{\^\i}tre}, {Leonardi}, {Leroy}, {Letendre}, {Levesque}, {Levin},
  {Leviton}, {Leyde}, {Li}, {Li}, {Li}, {Li}, {Li}, {Li}, {Lin}, {Lin}, {Lin},
  {Lin}, {Lin}, {Linde}, {Linker}, {Linley}, {Littenberg}, {Liu}, {Liu}, {Liu},
  {Liu}, {Llamas}, {Llorens-Monteagudo}, {Lo}, {Lockwood}, {Loh}, {London},
  {Longo}, {Lopez}, {Portilla}, {Lorenzini}, {Loriette}, {Lormand}, {Losurdo},
  {Lott}, {Lough}, {Lousto}, {Lovelace}, {Lucaccioni}, {L{\"u}ck}, {Lumaca},
  {Lundgren}, {Luo}, {Lynam}, {Macas}, {Macinnis}, {MacLeod}, {MacMillan},
  {Macquet}, {Hernandez}, {Magazz{\`u}}, {Magee}, {Maggiore}, {Magnozzi},
  {Mahesh}, {Majorana}, {Makarem}, {Maksimovic}, {Maliakal}, {Malik}, {Man},
  {Mandic}, {Mangano}, {Mango}, {Mansell}, {Manske}, {Mantovani}, {Mapelli},
  {Marchesoni}, {Marchio}, {Marion}, {Mark}, {M{\'a}rka}, {M{\'a}rka},
  {Markakis}, {Markosyan}, {Markowitz}, {Maros}, {Marquina}, {Marsat},
  {Martelli}, {Martin}, {Martin}, {Martinez}, {Martinez}, {Martinez},
  {Martinovic}, {Martynov}, {Marx}, {Masalehdan}, {Mason}, {Massera},
  {Masserot}, {Massinger}, {Masso-Reid}, {Mastrogiovanni}, {Matas},
  {Mateu-Lucena}, {Matichard}, {Matiushechkina}, {Mavalvala}, {McCann},
  {McCarthy}, {McClelland}, {McClincy}, {McCormick}, {McCuller}, {McGhee},
  {McGuire}, {McIsaac}, {McIver}, {McRae}, {McWilliams}, {Meacher}, {Mehmet},
  {Mehta}, {Meijer}, {Melatos}, {Melchor}, {Mendell}, {Menendez-Vazquez},
  {Menoni}, {Mercer}, {Mereni}, {Merfeld}, {Merilh}, {Merritt}, {Merzougui},
  {Meshkov}, {Messenger}, {Messick}, {Meyers}, {Meylahn}, {Mhaske}, {Miani},
  {Miao}, {Michaloliakos}, {Michel}, {Michimura}, {Middleton}, {Milano},
  {Miller}, {Miller}, {Miller}, {Miller}, {Millhouse}, {Mills}, {Milotti},
  {Minazzoli}, {Minenkov}, {Mio}, {Mir}, {Miravet-Ten{\'e}s}, {Mishra},
  {Mishra}, {Mistry}, {Mitra}, {Mitrofanov}, {Mitselmakher}, {Mittleman},
  {Miyakawa}, {Miyamoto}, {Miyazaki}, {Miyo}, {Miyoki}, {Mo}, {Modafferi},
  {Moguel}, {Mogushi}, {Mohapatra}, {Mohite}, {Molina}, {Molina-Ruiz},
  {Mondin}, {Montani}, {Moore}, {Moraru}, {Morawski}, {More}, {Moreno},
  {Moreno}, {Mori}, {Morisaki}, {Moriwaki}, {Morr{\'a}s}, {Mours}, {Mow-Lowry},
  {Mozzon}, {Muciaccia}, {Mukherjee}, {Mukherjee}, {Mukherjee}, {Mukherjee},
  {Mukherjee}, {Mukund}, {Mullavey}, {Munch}, {Mu{\~n}iz}, {Murray},
  {Musenich}, {Muusse}, {Nadji}, {Nagano}, {Nagano}, {Nagar}, {Nakamura},
  {Nakano}, {Nakano}, {Nakashima}, {Nakayama}, {Napolano}, {Nardecchia},
  {Narikawa}, {Naticchioni}, {Nayak}, {Nayak}, {Negishi}, {Neil}, {Neilson},
  {Nelemans}, {Nelson}, {Nery}, {Neubauer}, {Neunzert}, {Ng}, {Ng}, {Nguyen},
  {Nguyen}, {Nguyen}, {Quynh}, {Ni}, {Nichols}, {Nishizawa}, {Nissanke},
  {Nitoglia}, {Nocera}, {Norman}, {North}, {Nozaki}, {Siles}, {Nuttall},
  {Oberling}, {O'Brien}, {Obuchi}, {O'Dell}, {Oelker}, {Ogaki}, {Oganesyan},
  {Oh}, {Oh}, {Oh}, {Ohashi}, {Ohishi}, {Ohkawa}, {Ohme}, {Ohta}, {Okada},
  {Okutani}, {Okutomi}, {Olivetto}, {Oohara}, {Ooi}, {Oram}, {O'Reilly},
  {Ormiston}, {Ormsby}, {Ortega}, {O'Shaughnessy}, {O'Shea}, {Oshino},
  {Ossokine}, {Osthelder}, {Otabe}, {Ottaway}, {Overmier}, {Pace}, {Pagano},
  {Page}, {Pagliaroli}, {Pai}, {Pai}, {Palamos}, {Palashov}, {Palomba}, {Pan},
  {Pan}, {Panda}, {Pang}, {Pang}, {Pankow}, {Pannarale}, {Pant}, {Panther},
  {Paoletti}, {Paoli}, {Paolone}, {Parisi}, {Park}, {Park}, {Parker},
  {Pascucci}, {Pasqualetti}, {Passaquieti}, {Passuello}, {Patel}, {Pathak},
  {Patricelli}, {Patron}, {Paul}, {Payne}, {Pedraza}, {Pegoraro}, {Pele},
  {Arellano}, {Penn}, {Perego}, {Pereira}, {Pereira}, {Perez}, {P{\'e}rigois},
  {Perkins}, {Perreca}, {Perri{\`e}s}, {Petermann}, {Petterson}, {Pfeiffer},
  {Pham}, {Phukon}, {Piccinni}, {Pichot}, {Piendibene}, {Piergiovanni},
  {Pierini}, {Pierro}, {Pillant}, {Pillas}, {Pilo}, {Pinard}, {Pinto}, {Pinto},
  {Piotrzkowski}, {Piotrzkowski}, {Pirello}, {Pitkin}, {Placidi}, {Planas},
  {Plastino}, {Pluchar}, {Poggiani}, {Polini}, {Pong}, {Ponrathnam},
  {Popolizio}, {Porter}, {Poulton}, {Powell}, {Pracchia}, {Pradier},
  {Prajapati}, {Prasai}, {Prasanna}, {Pratten}, {Principe}, {Prodi},
  {Prokhorov}, {Prosposito}, {Prudenzi}, {Puecher}, {Punturo}, {Puosi},
  {Puppo}, {P{\"u}rrer}, {Qi}, {Quetschke}, {Quitzow-James}, {Raab},
  {Raaijmakers}, {Radkins}, {Radulesco}, {Raffai}, {Rail}, {Raja}, {Rajan},
  {Ramirez}, {Ramirez}, {Ramos-Buades}, {Rana}, {Rapagnani}, {Rapol}, {Ray},
  {Raymond}, {Raza}, {Razzano}, {Read}, {Rees}, {Regimbau}, {Rei}, {Reid},
  {Reid}, {Reitze}, {Relton}, {Renzini}, {Rettegno}, {Reza}, {Rezac}, {Ricci},
  {Richards}, {Richardson}, {Richardson}, {Riemenschneider}, {Riles},
  {Rinaldi}, {Rink}, {Rizzo}, {Robertson}, {Robie}, {Robinet}, {Rocchi},
  {Rodriguez}, {Rolland}, {Rollins}, {Romanelli}, {Romano}, {Romel},
  {Romero-Rodr{\'\i}guez}, {Romero-Shaw}, {Romie}, {Ronchini}, {Rosa}, {Rose},
  {Rosi{\'n}ska}, {Ross}, {Rowan}, {Rowlinson}, {Roy}, {Roy}, {Roy}, {Rozza},
  {Ruggi}, {Ryan}, {Sachdev}, {Sadecki}, {Sadiq}, {Sago}, {Saito}, {Saito},
  {Sakai}, {Sakai}, {Sakellariadou}, {Sakuno}, {Salafia}, {Salconi}, {Saleem},
  {Salemi}, {Samajdar}, {Sanchez}, {Sanchez}, {Sanchez}, {Sanchis-Gual},
  {Sanders}, {Sanuy}, {Saravanan}, {Sarin}, {Sassolas}, {Satari},
  {Sathyaprakash}, {Sato}, {Sato}, {Sauter}, {Savage}, {Sawada}, {Sawant},
  {Sawant}, {Sayah}, {Schaetzl}, {Scheel}, {Scheuer}, {Schiworski}, {Schmidt},
  {Schmidt}, {Schnabel}, {Schneewind}, {Schofield}, {Sch{\"o}nbeck}, {Schulte},
  {Schutz}, {Schwartz}, {Scott}, {Scott}, {Seglar-Arroyo}, {Sekiguchi},
  {Sekiguchi}, {Sellers}, {Sengupta}, {Sentenac}, {Seo}, {Sequino}, {Sergeev},
  {Setyawati}, {Shaffer}, {Shahriar}, {Shams}, {Shao}, {Sharma}, {Sharma},
  {Shawhan}, {Shcheblanov}, {Shibagaki}, {Shikauchi}, {Shimizu}, {Shimoda},
  {Shimode}, {Shinkai}, {Shishido}, {Shoda}, {Shoemaker}, {Shoemaker},
  {Shyamsundar}, {Sieniawska}, {Sigg}, {Singer}, {Singh}, {Singh}, {Singha},
  {Sintes}, {Sipala}, {Skliris}, {Slagmolen}, {Slaven-Blair}, {Smetana},
  {Smith}, {Smith}, {Soldateschi}, {Somala}, {Somiya}, {Son}, {Soni}, {Soni},
  {Sordini}, {Sorrentino}, {Sorrentino}, {Sotani}, {Soulard}, {Souradeep},
  {Sowell}, {Spagnuolo}, {Spencer}, {Spera}, {Srinivasan}, {Srivastava},
  {Srivastava}, {Staats}, {Stachie}, {Steer}, {Steinhoff}, {Steinlechner},
  {Steinlechner}, {Stevenson}, {Stops}, {Stover}, {Strain}, {Strang},
  {Stratta}, {Strunk}, {Sturani}, {Stuver}, {Sudhagar}, {Sudhir}, {Sugimoto},
  {Suh}, {Sullivan}, {Summerscales}, {Sun}, {Sun}, {Sunil}, {Sur}, {Suresh},
  {Sutton}, {Suzuki}, {Suzuki}, {Swinkels}, {Szczepa{\'n}czyk}, {Szewczyk},
  {Tacca}, {Tagoshi}, {Tait}, {Takahashi}, {Takahashi}, {Takamori}, {Takano},
  {Takeda}, {Takeda}, {Talbot}, {Talbot}, {Tanaka}, {Tanaka}, {Tanaka},
  {Tanaka}, {Tanaka}, {Tanasijczuk}, {Tanioka}, {Tanner}, {Tao}, {Tao},
  {Mart{\'\i}n}, {Taranto}, {Tasson}, {Telada}, {Tenorio}, {Terhune},
  {Terkowski}, {Thirugnanasambandam}, {Thomas}, {Thomas}, {Thomas}, {Thompson},
  {Thondapu}, {Thorne}, {Thrane}, {Tiwari}, {Tiwari}, {Tiwari}, {Toivonen},
  {Toland}, {Tolley}, {Tomaru}, {Tomigami}, {Tomura}, {Tonelli},
  {Torres-Forn{\'e}}, {Torrie}, {E Melo}, {T{\"o}yr{\"a}}, {Trapananti},
  {Travasso}, {Traylor}, {Trevor}, {Tringali}, {Tripathee}, {Troiano},
  {Trovato}, {Trozzo}, {Trudeau}, {Tsai}, {Tsai}, {Tsang}, {Tsang}, {Tsao},
  {Tse}, {Tso}, {Tsubono}, {Tsuchida}, {Tsukada}, {Tsuna}, {Tsutsui},
  {Tsuzuki}, {Turbang}, {Turconi}, {Tuyenbayev}, {Ubhi}, {Uchikata},
  {Uchiyama}, {Udall}, {Ueda}, {Uehara}, {Ueno}, {Ueshima}, {Unnikrishnan},
  {Uraguchi}, {Urban}, {Ushiba}, {Utina}, {Vahlbruch}, {Vajente}, {Vajpeyi},
  {Valdes}, {Valentini}, {Valsan}, {van Bakel}, {van Beuzekom}, {van den
  Brand}, {van den Broeck}, {Vander-Hyde}, {van der Schaaf}, {van Heijningen},
  {Vanosky}, {van Putten}, {van Remortel}, {Vardaro}, {Vargas}, {Varma},
  {Vas{\'u}th}, {Vecchio}, {Vedovato}, {Veitch}, {Veitch}, {Venneberg},
  {Venugopalan}, {Verkindt}, {Verma}, {Verma}, {Veske}, {Vetrano},
  {Vicer{\'e}}, {Vidyant}, {Viets}, {Vijaykumar}, {Villa-Ortega}, {Vinet},
  {Virtuoso}, {Vitale}, {Vo}, {Vocca}, {von Reis}, {von Wrangel}, {Vorvick},
  {Vyatchanin}, {Wade}, {Wade}, {Wagner}, {Walet}, {Walker}, {Wallace},
  {Wallace}, {Walsh}, {Wang}, {Wang}, {Wang}, {Ward}, {Warner}, {Was},
  {Washimi}, {Washington}, {Watchi}, {Weaver}, {Webster}, {Weinert},
  {Weinstein}, {Weiss}, {Weller}, {Wellmann}, {Wen}, {We{\ss}els}, {Wette},
  {Whelan}, {White}, {Whiting}, {Whittle}, {Wilken}, {Williams}, {Williams},
  {Williamson}, {Willis}, {Willke}, {Wilson}, {Winkler}, {Wipf}, {Wlodarczyk},
  {Woan}, {Woehler}, {Wofford}, {Wong}, {Wu}, {Wu}, {Wu}, {Wu}, {Wysocki},
  {Xiao}, {Xu}, {Yamada}, {Yamamoto}, {Yamamoto}, {Yamamoto}, {Yamamoto},
  {Yamashita}, {Yamazaki}, {Yang}, {Yang}, {Yang}, {Yang}, {Yang}, {Yap},
  {Yeeles}, {Yelikar}, {Ying}, {Yokogawa}, {Yokoyama}, {Yokozawa}, {Yoo},
  {Yoshioka}, {Yu}, {Yu}, {Yuzurihara}, {Zadro{\.z}ny}, {Zanolin}, {Zeidler},
  {Zelenova}, {Zendri}, {Zevin}, {Zhan}, {Zhang}, {Zhang}, {Zhang}, {Zhang},
  {Zhang}, {Zhao}, {Zhao}, {Zhao}, {Zhao}, {Zheng}, {Zhou}, {Zhou}, {Zhu},
  {Zhu}, {Zimmerman}, {Zlochower}, {Zucker}, {Zweizig}, {LIGO Scientific
  Collaboration}, {VIRGO Collaboration}, \& {KAGRA
  Collaboration}}]{2023PhRvX..13a1048A}
{Abbott}, R., {Abbott}, T.~D., {Acernese}, F., {et~al.} 2023, Physical Review
  X, 13, 011048

\bibitem[{{Alonso} {et~al.}(2020){Alonso}, {Contaldi}, {Cusin}, {Ferreira}, \&
  {Renzini}}]{2020PhRvD.101l4048A}
{Alonso}, D., {Contaldi}, C.~R., {Cusin}, G., {Ferreira}, P.~G., \& {Renzini},
  A.~I. 2020, \prd, 101, 124048

\bibitem[{{Amaro-Seoane} {et~al.}(2023){Amaro-Seoane}, {Andrews}, {Arca Sedda},
  {Askar}, {Baghi}, {Balasov}, {Bartos}, {Bavera}, {Bellovary}, {Berry},
  {Berti}, {Bianchi}, {Blecha}, {Blondin}, {Bogdanovi{\'c}}, {Boissier},
  {Bonetti}, {Bonoli}, {Bortolas}, {Breivik}, {Capelo}, {Caramete},
  {Cattorini}, {Charisi}, {Chaty}, {Chen}, {Chru{\'s}li{\'n}ska}, {Chua},
  {Church}, {Colpi}, {D'Orazio}, {Danielski}, {Davies}, {Dayal}, {De Rosa},
  {Derdzinski}, {Destounis}, {Dotti}, {Dutan}, {Dvorkin}, {Fabj}, {Foglizzo},
  {Ford}, {Fouvry}, {Franchini}, {Fragos}, {Fryer}, {Gaspari}, {Gerosa},
  {Graziani}, {Groot}, {Habouzit}, {Haggard}, {Haiman}, {Han}, {Istrate},
  {Johansson}, {Khan}, {Kimpson}, {Kokkotas}, {Kong}, {Korol}, {Kremer},
  {Kupfer}, {Lamberts}, {Larson}, {Lau}, {Liu}, {Lloyd-Ronning}, {Lodato},
  {Lupi}, {Ma}, {Maccarone}, {Mandel}, {Mangiagli}, {Mapelli}, {Mathis},
  {Mayer}, {McGee}, {McKernan}, {Miller}, {Mota}, {Mumpower}, {Nasim},
  {Nelemans}, {Noble}, {Pacucci}, {Panessa}, {Paschalidis}, {Pfister},
  {Porquet}, {Quenby}, {Ricarte}, {R{\"o}pke}, {Regan}, {Rosswog}, {Ruiter},
  {Ruiz}, {Runnoe}, {Schneider}, {Schnittman}, {Secunda}, {Sesana}, {Seto},
  {Shao}, {Shapiro}, {Sopuerta}, {Stone}, {Suvorov}, {Tamanini}, {Tamfal},
  {Tauris}, {Temmink}, {Tomsick}, {Toonen}, {Torres-Orjuela}, {Toscani},
  {Tsokaros}, {Unal}, {V{\'a}zquez-Aceves}, {Valiante}, {van Putten}, {van
  Roestel}, {Vignali}, {Volonteri}, {Wu}, {Younsi}, {Yu}, {Zane}, {Zwick},
  {Antonini}, {Baibhav}, {Barausse}, {Bonilla Rivera}, {Branchesi},
  {Branduardi-Raymont}, {Burdge}, {Chakraborty}, {Cuadra}, {Dage}, {Davis}, {de
  Mink}, {Decarli}, {Doneva}, {Escoffier}, {Gandhi}, {Haardt}, {Lousto},
  {Nissanke}, {Nordhaus}, {O'Shaughnessy}, {Portegies Zwart}, {Pound},
  {Schussler}, {Sergijenko}, {Spallicci}, {Vernieri}, \&
  {Vigna-G{\'o}mez}}]{2023LRR....26....2A}
{Amaro-Seoane}, P., {Andrews}, J., {Arca Sedda}, M., {et~al.} 2023, Living
  Reviews in Relativity, 26, 2

\bibitem[{{Amaro-Seoane} {et~al.}(2017){Amaro-Seoane}, {Audley}, {Babak},
  {Baker}, {Barausse}, {Bender}, {Berti}, {Binetruy}, {Born}, {Bortoluzzi},
  {Camp}, {Caprini}, {Cardoso}, {Colpi}, {Conklin}, {Cornish}, {Cutler},
  {Danzmann}, {Dolesi}, {Ferraioli}, {Ferroni}, {Fitzsimons}, {Gair}, {Gesa
  Bote}, {Giardini}, {Gibert}, {Grimani}, {Halloin}, {Heinzel}, {Hertog},
  {Hewitson}, {Holley-Bockelmann}, {Hollington}, {Hueller}, {Inchauspe},
  {Jetzer}, {Karnesis}, {Killow}, {Klein}, {Klipstein}, {Korsakova}, {Larson},
  {Livas}, {Lloro}, {Man}, {Mance}, {Martino}, {Mateos}, {McKenzie},
  {McWilliams}, {Miller}, {Mueller}, {Nardini}, {Nelemans}, {Nofrarias},
  {Petiteau}, {Pivato}, {Plagnol}, {Porter}, {Reiche}, {Robertson},
  {Robertson}, {Rossi}, {Russano}, {Schutz}, {Sesana}, {Shoemaker}, {Slutsky},
  {Sopuerta}, {Sumner}, {Tamanini}, {Thorpe}, {Troebs}, {Vallisneri},
  {Vecchio}, {Vetrugno}, {Vitale}, {Volonteri}, {Wanner}, {Ward}, {Wass},
  {Weber}, {Ziemer}, \& {Zweifel}}]{2017arXiv170200786A}
{Amaro-Seoane}, P., {Audley}, H., {Babak}, S., {et~al.} 2017, arXiv e-prints,
  arXiv:1702.00786

\bibitem[{{Auclair} {et~al.}(2023){Auclair}, {Bacon}, {Baker}, {Barreiro},
  {Bartolo}, {Belgacem}, {Bellomo}, {Ben-Dayan}, {Bertacca}, {Besancon},
  {Blanco-Pillado}, {Blas}, {Boileau}, {Calcagni}, {Caldwell}, {Caprini},
  {Carbone}, {Chang}, {Chen}, {Christensen}, {Clesse}, {Comelli}, {Congedo},
  {Contaldi}, {Crisostomi}, {Croon}, {Cui}, {Cusin}, {Cutting}, {Dalang}, {De
  Luca}, {Pozzo}, {Desjacques}, {Dimastrogiovanni}, {Dorsch}, {Ezquiaga},
  {Fasiello}, {Figueroa}, {Flauger}, {Franciolini}, {Frusciante}, {Fumagalli},
  {Garc{\'\i}a-Bellido}, {Gould}, {Holz}, {Iacconi}, {Jain}, {Jenkins},
  {Jinno}, {Joana}, {Karnesis}, {Konstandin}, {Koyama}, {Kozaczuk},
  {Kuroyanagi}, {Laghi}, {Lewicki}, {Lombriser}, {Madge}, {Maggiore},
  {Malhotra}, {Mancarella}, {Mandic}, {Mangiagli}, {Matarrese}, {Mazumdar},
  {Mukherjee}, {Musco}, {Nardini}, {No}, {Papanikolaou}, {Peloso}, {Pieroni},
  {Pilo}, {Raccanelli}, {Renaux-Petel}, {Renzini}, {Ricciardone}, {Riotto},
  {Romano}, {Rollo}, {Pol}, {Morales}, {Sakellariadou}, {Saltas}, {Scalisi},
  {Schmitz}, {Schwaller}, {Sergijenko}, {Servant}, {Simakachorn}, {Sorbo},
  {Sousa}, {Speri}, {Steer}, {Tamanini}, {Tasinato}, {Torrado}, {Unal},
  {Vennin}, {Vernieri}, {Vernizzi}, {Volonteri}, {Wachter}, {Wands},
  {Witkowski}, {Zumalac{\'a}rregui}, {Annis}, {Ares}, {Avelino}, {Avgoustidis},
  {Barausse}, {Bonilla}, {Bonvin}, {Bosso}, {Calabrese},
  {{\c{c}}al{\i}{\c{s}}kan}, {Cembranos}, {Chala}, {Chernoff}, {Clough},
  {Criswell}, {Das}, {Silva}, {Dayal}, {Domcke}, {Durrer}, {Easther},
  {Escoffier}, {Ferrans}, {Fryer}, {Gair}, {Gordon}, {Hendry}, {Hindmarsh},
  {Hooper}, {Kajfasz}, {Kopp}, {Koushiappas}, {Kumar}, {Kunz}, {Lagos},
  {Lilley}, {Lizarraga}, {Lobo}, {Maleknejad}, {Martins}, {Meerburg}, {Meyer},
  {Mimoso}, {Nesseris}, {Nunes}, {Oikonomou}, {Orlando}, {{\"O}zsoy},
  {Pacucci}, {Palmese}, {Petiteau}, {Pinol}, {Zwart}, {Pratten}, {Prokopec},
  {Quenby}, {Rastgoo}, {Roest}, {Rummukainen}, {Schimd}, {Secroun}, {Sesana},
  {Sopuerta}, {Tereno}, {Tolley}, {Urrestilla}, {Vagenas}, {van de Vis}, {van
  de Weygaert}, {Wardell}, {Weir}, {White}, {{\'S}wie{\.Z}ewska}, {Zhdanov}, \&
  {LISA Cosmology Working Group}}]{2023LRR....26....5A}
{Auclair}, P., {Bacon}, D., {Baker}, T., {et~al.} 2023, Living Reviews in
  Relativity, 26, 5

\bibitem[{{Babak} {et~al.}(2023){Babak}, {Caprini}, {Figueroa}, {Karnesis},
  {Marcoccia}, {Nardini}, {Pieroni}, {Ricciardone}, {Sesana}, \&
  {Torrado}}]{2023JCAP...08..034B}
{Babak}, S., {Caprini}, C., {Figueroa}, D.~G., {et~al.} 2023, \jcap, 2023, 034

\bibitem[{{Badenes} {et~al.}(2018){Badenes}, {Mazzola}, {Thompson}, {Covey},
  {Freeman}, {Walker}, {Moe}, {Troup}, {Nidever}, {Allende Prieto}, {Andrews},
  {Barb{\'a}}, {Beers}, {Bovy}, {Carlberg}, {De Lee}, {Johnson}, {Lewis},
  {Majewski}, {Pinsonneault}, {Sobeck}, {Stassun}, {Stringfellow}, \&
  {Zasowski}}]{2018ApJ...854..147B}
{Badenes}, C., {Mazzola}, C., {Thompson}, T.~A., {et~al.} 2018, \apj, 854, 147

\bibitem[{{Baghi}(2022)}]{2022arXiv220412142B}
{Baghi}, Q. 2022, arXiv e-prints, arXiv:2204.12142

\bibitem[{{Bavera} {et~al.}(2022){Bavera}, {Franciolini}, {Cusin}, {Riotto},
  {Zevin}, \& {Fragos}}]{2022A&A...660A..26B}
{Bavera}, S.~S., {Franciolini}, G., {Cusin}, G., {et~al.} 2022, \aap, 660, A26

\bibitem[{{Bin{\'e}truy} {et~al.}(2012){Bin{\'e}truy}, {Boh{\'e}}, {Caprini},
  \& {Dufaux}}]{2012JCAP...06..027B}
{Bin{\'e}truy}, P., {Boh{\'e}}, A., {Caprini}, C., \& {Dufaux}, J.-F. 2012,
  \jcap, 2012, 027

\bibitem[{{Caprini} {et~al.}(2009){Caprini}, {Durrer}, \&
  {Servant}}]{2009JCAP...12..024C}
{Caprini}, C., {Durrer}, R., \& {Servant}, G. 2009, \jcap, 2009, 024

\bibitem[{{Caprini} {et~al.}(2016){Caprini}, {Hindmarsh}, {Huber},
  {Konstandin}, {Kozaczuk}, {Nardini}, {No}, {Petiteau}, {Schwaller},
  {Servant}, \& {Weir}}]{2016JCAP...04..001C}
{Caprini}, C., {Hindmarsh}, M., {Huber}, S., {et~al.} 2016, \jcap, 2016, 001

\bibitem[{{Christensen}(2019)}]{2019RPPh...82a6903C}
{Christensen}, N. 2019, Reports on Progress in Physics, 82, 016903

\bibitem[{{Chruslinska} \& {Nelemans}(2019)}]{2019MNRAS.488.5300C}
{Chruslinska}, M. \& {Nelemans}, G. 2019, \mnras, 488, 5300

\bibitem[{{Cusin} {et~al.}(2020){Cusin}, {Dvorkin}, {Pitrou}, \&
  {Uzan}}]{2020MNRAS.493L...1C}
{Cusin}, G., {Dvorkin}, I., {Pitrou}, C., \& {Uzan}, J.-P. 2020, \mnras, 493,
  L1

\bibitem[{{Dvorkin} {et~al.}(2016){Dvorkin}, {Vangioni}, {Silk}, {Uzan}, \&
  {Olive}}]{2016MNRAS.461.3877D}
{Dvorkin}, I., {Vangioni}, E., {Silk}, J., {Uzan}, J.-P., \& {Olive}, K.~A.
  2016, \mnras, 461, 3877

\bibitem[{{Edlund} {et~al.}(2005){Edlund}, {Tinto}, {Kr{\'o}lak}, \&
  {Nelemans}}]{2005PhRvD..71l2003E}
{Edlund}, J.~A., {Tinto}, M., {Kr{\'o}lak}, A., \& {Nelemans}, G. 2005, \prd,
  71, 122003

\bibitem[{{Farmer} \& {Phinney}(2003)}]{2003MNRAS.346.1197F}
{Farmer}, A.~J. \& {Phinney}, E.~S. 2003, \mnras, 346, 1197

\bibitem[{Hawking \& Israel(1987)}]{hawking1987three}
Hawking, S.~W. \& Israel, W. 1987, Three hundred years of gravitation
  (Cambridge University Press)

\bibitem[{{Karnesis} {et~al.}(2021){Karnesis}, {Babak}, {Pieroni}, {Cornish},
  \& {Littenberg}}]{2021PhRvD.104d3019K}
{Karnesis}, N., {Babak}, S., {Pieroni}, M., {Cornish}, N., \& {Littenberg}, T.
  2021, \prd, 104, 043019

\bibitem[{{Keim} {et~al.}(2023){Keim}, {Korol}, \&
  {Rossi}}]{2023MNRAS.521.1088K}
{Keim}, M.~A., {Korol}, V., \& {Rossi}, E.~M. 2023, \mnras, 521, 1088

\bibitem[{{Korol} {et~al.}(2020){Korol}, {Toonen}, {Klein}, {Belokurov},
  {Vincenzo}, {Buscicchio}, {Gerosa}, {Moore}, {Roebber}, {Rossi}, \&
  {Vecchio}}]{2020A&A...638A.153K}
{Korol}, V., {Toonen}, S., {Klein}, A., {et~al.} 2020, \aap, 638, A153

\bibitem[{{Kowalska-Leszczynska} {et~al.}(2015){Kowalska-Leszczynska},
  {Regimbau}, {Bulik}, {Dominik}, \& {Belczynski}}]{2015A&A...574A..58K}
{Kowalska-Leszczynska}, I., {Regimbau}, T., {Bulik}, T., {Dominik}, M., \&
  {Belczynski}, K. 2015, \aap, 574, A58

\bibitem[{{Kroupa}(2001)}]{2001MNRAS.322..231K}
{Kroupa}, P. 2001, \mnras, 322, 231

\bibitem[{{Lehoucq} {et~al.}(2023){Lehoucq}, {Dvorkin}, {Srinivasan},
  {Pellouin}, \& {Lamberts}}]{2023MNRAS.tmp.2801L}
{Lehoucq}, L., {Dvorkin}, I., {Srinivasan}, R., {Pellouin}, C., \& {Lamberts},
  A. 2023, \mnras [\eprint[arXiv]{2306.09861}]

\bibitem[{{Luo} {et~al.}(2016){Luo}, {Chen}, {Duan}, {Gong}, {Hu}, {Ji}, {Liu},
  {Mei}, {Milyukov}, {Sazhin}, {Shao}, {Toth}, {Tu}, {Wang}, {Wang}, {Yeh},
  {Zhan}, {Zhang}, {Zharov}, \& {Zhou}}]{2016CQGra..33c5010L}
{Luo}, J., {Chen}, L.-S., {Duan}, H.-Z., {et~al.} 2016, Classical and Quantum
  Gravity, 33, 035010

\bibitem[{{Luo} {et~al.}(2021){Luo}, {Wang}, {Wu}, {Hu}, \&
  {Jin}}]{2021PTEP.2021eA108L}
{Luo}, Z., {Wang}, Y., {Wu}, Y., {Hu}, W., \& {Jin}, G. 2021, Progress of
  Theoretical and Experimental Physics, 2021, 05A108

\bibitem[{Madau \& Dickinson(2014)}]{madau_cosmic_2014}
Madau, P. \& Dickinson, M. 2014, Annual Review of Astronomy and Astrophysics,
  52, 415, \_eprint: https://doi.org/10.1146/annurev-astro-081811-125615

\bibitem[{{Moe} \& {Di Stefano}(2017)}]{2017ApJS..230...15M}
{Moe}, M. \& {Di Stefano}, R. 2017, \apjs, 230, 15

\bibitem[{{Moe} {et~al.}(2019){Moe}, {Kratter}, \&
  {Badenes}}]{2019ApJ...875...61M}
{Moe}, M., {Kratter}, K.~M., \& {Badenes}, C. 2019, \apj, 875, 61

\bibitem[{{Nelemans} {et~al.}(2001{\natexlab{a}}){Nelemans}, {Yungelson}, \&
  {Portegies Zwart}}]{2001A&A...375..890N}
{Nelemans}, G., {Yungelson}, L.~R., \& {Portegies Zwart}, S.~F.
  2001{\natexlab{a}}, \aap, 375, 890

\bibitem[{{Nelemans} {et~al.}(2004){Nelemans}, {Yungelson}, \& {Portegies
  Zwart}}]{2004MNRAS.349..181N}
{Nelemans}, G., {Yungelson}, L.~R., \& {Portegies Zwart}, S.~F. 2004, \mnras,
  349, 181

\bibitem[{{Nelemans} {et~al.}(2001{\natexlab{b}}){Nelemans}, {Yungelson},
  {Portegies Zwart}, \& {Verbunt}}]{2001A&A...365..491N}
{Nelemans}, G., {Yungelson}, L.~R., {Portegies Zwart}, S.~F., \& {Verbunt}, F.
  2001{\natexlab{b}}, \aap, 365, 491

\bibitem[{Peters \& Mathews(1963)}]{peters_gravitational_1963}
Peters, P.~C. \& Mathews, J. 1963, Physical Review, 131, 435

\bibitem[{{Phinney}(2001)}]{2001astro.ph..8028P}
{Phinney}, E.~S. 2001, arXiv e-prints, astro

\bibitem[{Phinney(2001)}]{phinney_practical_2001}
Phinney, E.~S. 2001, A {Practical} {Theorem} on {Gravitational} {Wave}
  {Backgrounds}, arXiv:astro-ph/0108028

\bibitem[{{Planck Collaboration} {et~al.}(2020){Planck Collaboration}, Aghanim,
  Akrami, Ashdown, Aumont, Baccigalupi, Ballardini, Banday, Barreiro, Bartolo,
  Basak, Battye, Benabed, Bernard, Bersanelli, Bielewicz, Bock, Bond, Borrill,
  Bouchet, Boulanger, Bucher, Burigana, Butler, Calabrese, Cardoso, Carron,
  Challinor, Chiang, Chluba, Colombo, Combet, Contreras, Crill, Cuttaia,
  De~Bernardis, De~Zotti, Delabrouille, Delouis, Di~Valentino, Diego, Doré,
  Douspis, Ducout, Dupac, Dusini, Efstathiou, Elsner, Enßlin, Eriksen,
  Fantaye, Farhang, Fergusson, Fernandez-Cobos, Finelli, Forastieri, Frailis,
  Fraisse, Franceschi, Frolov, Galeotta, Galli, Ganga, Génova-Santos, Gerbino,
  Ghosh, González-Nuevo, Górski, Gratton, Gruppuso, Gudmundsson, Hamann,
  Handley, Hansen, Herranz, Hildebrandt, Hivon, Huang, Jaffe, Jones, Karakci,
  Keihänen, Keskitalo, Kiiveri, Kim, Kisner, Knox, Krachmalnicoff, Kunz,
  Kurki-Suonio, Lagache, Lamarre, Lasenby, Lattanzi, Lawrence, Le~Jeune, Lemos,
  Lesgourgues, Levrier, Lewis, Liguori, Lilje, Lilley, Lindholm,
  López-Caniego, Lubin, Ma, Macías-Pérez, Maggio, Maino, Mandolesi,
  Mangilli, Marcos-Caballero, Maris, Martin, Martinelli, Martínez-González,
  Matarrese, Mauri, McEwen, Meinhold, Melchiorri, Mennella, Migliaccio, Millea,
  Mitra, Miville-Deschênes, Molinari, Montier, Morgante, Moss, Natoli,
  Nørgaard-Nielsen, Pagano, Paoletti, Partridge, Patanchon, Peiris, Perrotta,
  Pettorino, Piacentini, Polastri, Polenta, Puget, Rachen, Reinecke,
  Remazeilles, Renzi, Rocha, Rosset, Roudier, Rubiño-Martín, Ruiz-Granados,
  Salvati, Sandri, Savelainen, Scott, Shellard, Sirignano, Sirri, Spencer,
  Sunyaev, Suur-Uski, Tauber, Tavagnacco, Tenti, Toffolatti, Tomasi, Trombetti,
  Valenziano, Valiviita, Van~Tent, Vibert, Vielva, Villa, Vittorio, Wandelt,
  Wehus, White, White, Zacchei, \& Zonca}]{planck_collaboration_planck_2020}
{Planck Collaboration}, Aghanim, N., Akrami, Y., {et~al.} 2020, Astronomy \&
  Astrophysics, 641, A6

\bibitem[{{Portegies Zwart} \& {Verbunt}(1996)}]{1996A&A...309..179P}
{Portegies Zwart}, S.~F. \& {Verbunt}, F. 1996, \aap, 309, 179

\bibitem[{{Punturo} {et~al.}(2010){Punturo}, {Abernathy}, {Acernese}, {Allen},
  {Andersson}, {Arun}, {Barone}, {Barr}, {Barsuglia}, {Beker}, {Beveridge},
  {Birindelli}, {Bose}, {Bosi}, {Braccini}, {Bradaschia}, {Bulik}, {Calloni},
  {Cella}, {Chassande Mottin}, {Chelkowski}, {Chincarini}, {Clark}, {Coccia},
  {Colacino}, {Colas}, {Cumming}, {Cunningham}, {Cuoco}, {Danilishin},
  {Danzmann}, {De Luca}, {De Salvo}, {Dent}, {De Rosa}, {Di Fiore}, {Di
  Virgilio}, {Doets}, {Fafone}, {Falferi}, {Flaminio}, {Franc}, {Frasconi},
  {Freise}, {Fulda}, {Gair}, {Gemme}, {Gennai}, {Giazotto}, {Glampedakis},
  {Granata}, {Grote}, {Guidi}, {Hammond}, {Hannam}, {Harms}, {Heinert},
  {Hendry}, {Heng}, {Hennes}, {Hild}, {Hough}, {Husa}, {Huttner}, {Jones},
  {Khalili}, {Kokeyama}, {Kokkotas}, {Krishnan}, {Lorenzini}, {L{\"u}ck},
  {Majorana}, {Mandel}, {Mandic}, {Martin}, {Michel}, {Minenkov}, {Morgado},
  {Mosca}, {Mours}, {M{\"u}ller{\textendash}Ebhardt}, {Murray}, {Nawrodt},
  {Nelson}, {Oshaughnessy}, {Ott}, {Palomba}, {Paoli}, {Parguez},
  {Pasqualetti}, {Passaquieti}, {Passuello}, {Pinard}, {Poggiani}, {Popolizio},
  {Prato}, {Puppo}, {Rabeling}, {Rapagnani}, {Read}, {Regimbau}, {Rehbein},
  {Reid}, {Rezzolla}, {Ricci}, {Richard}, {Rocchi}, {Rowan}, {R{\"u}diger},
  {Sassolas}, {Sathyaprakash}, {Schnabel}, {Schwarz}, {Seidel}, {Sintes},
  {Somiya}, {Speirits}, {Strain}, {Strigin}, {Sutton}, {Tarabrin},
  {Th{\"u}ring}, {van den Brand}, {van Leewen}, {van Veggel}, {van den Broeck},
  {Vecchio}, {Veitch}, {Vetrano}, {Vicere}, {Vyatchanin}, {Willke}, {Woan},
  {Wolfango}, \& {Yamamoto}}]{2010CQGra..27s4002P}
{Punturo}, M., {Abernathy}, M., {Acernese}, F., {et~al.} 2010, Classical and
  Quantum Gravity, 27, 194002

\bibitem[{{Reitze} {et~al.}(2019){Reitze}, {Adhikari}, {Ballmer}, {Barish},
  {Barsotti}, {Billingsley}, {Brown}, {Chen}, {Coyne}, {Eisenstein}, {Evans},
  {Fritschel}, {Hall}, {Lazzarini}, {Lovelace}, {Read}, {Sathyaprakash},
  {Shoemaker}, {Smith}, {Torrie}, {Vitale}, {Weiss}, {Wipf}, \&
  {Zucker}}]{2019BAAS...51g..35R}
{Reitze}, D., {Adhikari}, R.~X., {Ballmer}, S., {et~al.} 2019, in Bulletin of
  the American Astronomical Society, Vol.~51, 35

\bibitem[{{Renzini} {et~al.}(2022){Renzini}, {Goncharov}, {Jenkins}, \&
  {Meyers}}]{renzini_stochastic_2022}
{Renzini}, A.~I., {Goncharov}, B., {Jenkins}, A.~C., \& {Meyers}, P.~M. 2022,
  Galaxies, 10, 34

\bibitem[{{Schneider} {et~al.}(2001){Schneider}, {Ferrari}, {Matarrese}, \&
  {Portegies Zwart}}]{2001MNRAS.324..797S}
{Schneider}, R., {Ferrari}, V., {Matarrese}, S., \& {Portegies Zwart}, S.~F.
  2001, \mnras, 324, 797

\bibitem[{{Schneider} {et~al.}(2010){Schneider}, {Marassi}, \&
  {Ferrari}}]{2010CQGra..27s4007S}
{Schneider}, R., {Marassi}, S., \& {Ferrari}, V. 2010, Classical and Quantum
  Gravity, 27, 194007

\bibitem[{{Thrane} \& {Romano}(2013)}]{2013PhRvD..88l4032T}
{Thrane}, E. \& {Romano}, J.~D. 2013, \prd, 88, 124032

\bibitem[{{Toonen} {et~al.}(2012){Toonen}, {Nelemans}, \& {Portegies
  Zwart}}]{2012A&A...546A..70T}
{Toonen}, S., {Nelemans}, G., \& {Portegies Zwart}, S. 2012, \aap, 546, A70

\bibitem[{{van Oirschot} {et~al.}(2014){van Oirschot}, {Nelemans}, {Toonen},
  {Pols}, {Brown}, {Helmi}, \& {Portegies Zwart}}]{2014A&A...569A..42V}
{van Oirschot}, P., {Nelemans}, G., {Toonen}, S., {et~al.} 2014, \aap, 569, A42

\end{thebibliography}
%

\end{document}